\documentclass[epj]{svjour}
\usepackage{newtxtext,newtxmath}
\usepackage{amsmath}
\usepackage{tabularx} 
\usepackage{graphicx}      
\usepackage{float}         
\usepackage{caption}       
\usepackage{subcaption}    
\usepackage{wrapfig}  
\usepackage{hyperref} 
 \usepackage{xcolor}

\usepackage[T1]{fontenc}
\begin{document}
\title{Prospective constraints on dark energy from nanohertz individual gravitational wave sources}
\author{Qing Yang\inst{1,2} \and Gu-yue Zhang\inst{1} \and Yi Huang\inst{1} \and Xiao Guo\inst{3,4}%
\thanks{\emph{Present address:} guoxiao@nao.cas.cn}%
}                     
\offprints{}          
\institute{College of Engineering Physics, Shenzhen Technology University, No.3002 Lantian Road, Shenzhen 518118, China \and Shenzhen Key Laboratory of Ultraintense Laser and Advanced Material Technology, Shenzhen Technology University, No.3002 Lantian Road, Shenzhen 518118, China \and
Institute for Gravitational Wave Astronomy, Henan Academy of Sciences, Zhengzhou 450046, Henan, China \and School of Fundamental Physics and Mathematical Sciences, Hangzhou Institute for Advanced Study, University of Chinese Academy of Sciences, No.1 Xiangshan Branch, Hangzhou 310024, China}
\date{Received: date / Revised version: date}
%
\abstract{
	Nanohertz gravitational waves (GWs) from supermassive binary black holes (SMBBHs), detectable via pulsar timing arrays (PTAs), offer a novel avenue to constrain dark energy. Based on cosmological simulations and semi-analytic galaxy formation models, this study explores the detectability of individual nanohertz SMBBH sources using next-generation PTAs and their potential for constraining dark energy under an optimistic scenario considering only the presence of white noise. By constructing light-cone SMBBH populations across hardening timescales ($\tau_H = 0.1/5/10$Gyr) and computing signal-to-noise ratios (SNR), we find advanced PTAs can resolve $10^2$--$10^3$ sources with SNR $> 8$ (primarily at $z < 1$ with chirp masses of $10^8$--$10^{10}M_{\odot}$). If electromagnetic counterparts can be identified, optimal configurations ($\sigma_t = 50$ns, $N_p = 1000$, $T_{\text{obs}} = 30$yr with$ \tau_H \leq 5$Gyr) could constrain the dark energy equation-of-state (EoS) parameter $w$ to $\Delta w \sim 0.023$--$0.048$, where the constraints only exhibit weak dependence on $\tau_H$ within $0.1$--$5$Gyr. If only $10\%$ of GW sources have detectable electromagnetic counterparts, constraints weaken to $\Delta w = 0.075$ ($\tau_H = 0.1$Gyr) and $\Delta w = 0.162$ ($\tau_H = 5$Gyr) under the most optimal parameter configuration. What's more, conservative PTAs ($N_p = 500$, $\sigma_t = 100$--$200$ns) with additional $30$-year data accumulation could double resolvable source counts and improve $\Delta w$ precision by $\sim 40\%$.
%
%
}

\titlerunning{Constraining Dark Energy with nHz GW Sources}
\maketitle
\section{Introduction}
Precise measurements of cosmological parameters play a pivotal role in understanding the universe's composition and evolution, especially in probing the nature of dark energy. Although various observational techniques have been developed, such as cosmic microwave background (CMB, \cite{Planck:2018vyg}) analysis, baryon acoustic oscillations (BAO, \cite{eBOSS:2020yzd,DES:2021wwk}), and type Ia supernovae (SN Ia, \cite{Riess:2019cxk}), persistent discrepancies between different methods highlight the need for independent verification. For instance, the 4.4$\sigma$ tension between local $H_0$ measurements from SN Ia and CMB-derived values might indicate new physics or unconsidered systematic errors \cite{Planck:2018vyg,Riess:2019cxk,DiValentino:2021izs,Abdalla:2022yfr}. On the other hand, the Dark Energy Spectroscopic Instrument (DESI) collaboration has presented preliminary constraints on dark energy evolution from the combination of its first-year data and CMB or SN Ia recently, reporting intriguing deviations from the standard cosmological model at low redshifts \cite{DESI:2024mwx}. These results, while promising, may also require independent verification.

Gravitational-wave astronomy has undergone transformative advancements since the first direct detection of gravitationla wave (GW) signal in 2015 \cite{LIGOScientific:2016aoc}, marking the dawn of a new observational era \cite{LIGOScientific:2021yby}. This breakthrough was followed by the landmark observation of a binary neutron star merger (GW170817, \cite{LIGOScientific:2017vwq}), which produced simultaneous gravitational and electromagnetic signals, enabling multi-messenger studies that resolved long-standing questions about heavy element nucleosynthesis \cite{Drout:2017ijr}. What's more, this particular event has pioneered a novel cosmological probe known as the ``standard siren" \cite{Schutz:1986gp,Finn:1992xs,Finn:1995ah,LIGOScientific:2017adf,Hotokezaka:2018dfi}, which may circumvents traditional cosmic distance ladder calibrations, as the gravitational waveform directly encodes the source's luminosity distance while optical observations of the associated kilonova provide the host galaxy's redshift. It is expected that this method would eliminates systematic uncertainties tied to traditional methods. Other related works utilizing gravitational wave signals for cosmological probes can be found in: \cite{Wang:2020xwn,LIGOScientific:2021aug,Yang:2022uye,Yu:2023ico,Palmese:2023beh,Borghi:2023opd,Pierra:2023deu,Mancarella:2024qle,Hanselman:2024hqy,Zheng:2024mbo}. At present, though, ground-based detectors predominantly observe stellar-mass black hole mergers which exhibit a lower probability of producing electromagnetic counterparts, thus partially limiting their application in cosmology.

This work mainly explores GW sources within the nanohertz frequency band: supermassive binary black holes (SMBBHs) with masses at the order of $10^6 M_\odot$ to $10^{10} M_\odot$\cite{2013ARA&A..51..511K}. Predicted to form via galactic mergers, these systems represent prime targets for pulsar timing arrays (PTAs)\cite{1978SvA....22...36S,1979ApJ...234.1100D,1984JApA....5..369B,1990ApJ...361..300F,2009MNRAS.394.2255S,2010CQGra..27h4016S,2011MNRAS.414.3251L,manchester2013pulsar,2013CQGra..30x4009S,2014MNRAS.444.3709Z,2016MNRAS.459.1737S,2019BAAS...51c.336T,2019MNRAS.483..503Y,Chen:2020qlp,2022ApJ...939...55G,2025ApJ...978..104G}. The global PTA network currently includes the Parkes PTA (PPTA; \cite{manchester2013pulsar,2023ApJ...951L...6R}), European PTA (EPTA, \cite{2013CQGra..30v4009K,2023A&A...678A..50E}), North American Nanohertz Observatory for Gravitational Waves (NANOGrav, \cite{2013CQGra..30v4008M,2019BAAS...51g.195R,2023ApJ...951L...8A}), Indian Pulsar Timing Array (InPTA, \cite{2018JApA...39...51J,2023A&A...678A..50E}), Chinese PTA (CPTA; \cite{2011IJMPD..20..989N,2009A&A...505..919S,2023RAA....23g5024X}), and MeerKAT PTA (MPTA; \cite{2023MNRAS.519.3976M}). Among these collaborations, PPTA, EPTA, and NANOGrav have accumulated over a decade of timing data for GW searches. These groups further synergize through the International PTA (IPTA, \cite{2013CQGra..30v4010M,10.1093/mnras/stw347,2019MNRAS.490.4666P}), which harmonizes data sharing and joint analyses. Future capabilities will be significantly enhanced by the Square Kilometer Array (SKA) \cite{Lazio2013SKA,2017PhRvL.118o1104W}, projected to discover thousands of millisecond pulsars and establish the ultra-sensitive SKA-PTA.
Recent breakthroughs from CPTA \cite{2023RAA....23g5024X}, NANOGrav \cite{2023ApJ...951L...8A}, EPTA+InPTA \cite{2023A&A...678A..50E}, and PPTA \cite{2023ApJ...951L...6R} have jointly reported evidence for a nanohertz GWB at 2-4$\sigma$ confidence (see also \cite{2020ApJ...905L..34A}). While the origin of this signal remains debated, current interpretations favor contributions from inspiralling SMBBHs \cite{2023ApJ...955..132C,2023ApJ...952L..37A,2023arXiv230616227A}, suggesting imminent prospects for resolving individual SMBBH sources.

Current PTA efforts have focused on stochastic GW backgrounds, but technological advances are pushing detection thresholds toward resolving individual SMBBHs. Recent simulations suggest that SKA-era PTAs could detect individual SMBBHs with high signal-to-noise ratios (SNR) 
\cite{Yang:2024mqz,2023ApJ...955..132C,Truant2025LightingUT}. The matter-rich environments of merging galaxies hosting supermassive black holes (SMBHs) substantially enhance the detectability of electromagnetic counterparts associated with these compact objects. When combined with redshift information from electromagnetic surveys, these systems could serve as standard sirens complementing existing probes, and the direct luminosity distance measurements from GW waveforms could be used to offer constraints on dark energy equation of state (EoS) parameters \cite{Yan_2020,Jin:2023zhi,Wang:2022llq}. 

In this work, we intend to employ the variants proposed by \cite{2011MNRAS.413..101G,2013MNRAS.428.1351G} of the semi-analytic galaxy formation model (SAM) grounded on dark matter (DM) halo merger trees derived from the Millennium simulation \cite{2005Natur.435..629S,2011MNRAS.413..101G} to construct multiple realizations of a comprehensive dataset of SMBBHs within the mock observable universe under different hardening timescales. From which we compute the SNR for each SMBBH event in all the realizations across diverse PTA parameter configurations (observational duration $T_{\rm obs}$, pulsar numbers $N_p$, and timing noise $\sigma_t$). Through systematic selecting of resolvable individual sources in all realizations under each parameter set, we subsequently employ these selected populations within the Fisher matrix formalism to constrain the dark energy EoS parameter $w$, which in this paper we assume to be a constant.

The paper is organized as follows: In Section \ref{sec:Fisher} we introduce the Fisher matrix framework for constraining dark energy parameters with PTA band individual GW events. In Section~\ref{sec:dataset}, we describe the galaxy evolution model and mock data we used, as well as the basic setups and methods for constructing light-cone SMBBHs in the observable universe and selecting bright individual sources under different PTA parameter configurations. In Section~\ref{sec:results}, we present our main results on the number, distribution, and other properties of the resolvable individual GW sources identified through our selection criteria. Additionally, we quantify the measurement precision for dark energy EoS parameter achievable with these sources, as derived from Fisher matrix analysis. Finally, Section~\ref{sec:concl} is devoted to conlusions and discussions. Throughout this work, we adopt a flat $\Lambda$CDM cosmology with $\Omega_m = 0.272$, $\Omega_\Lambda = 0.728$, and $H_0 = 70.4$ km s$^{-1}$ Mpc$^{-1}$.
\section{Fisher Matrix Methodology for EoM of Dark Energy}
\label{sec:Fisher}

PTAs utilize observations of millisecond pulsars (MSPs) to detect nanoHertz gravitational waves. The measured times of arrival (ToAs) of radio pulses from MSPs—monitored with a cadence ranging from bi-weekly to monthly over observational timescales spanning decades—are compared against theoretical predictions. The deviations between observed and modeled ToAs, known as ToA residuals, encode potential gravitational wave signals. By analyzing these timing residuals, researchers can both investigate potential gravitational wave signals and utilize the cosmological information embedded within them to quantify the measurement precision of relevant cosmological parameters. We will focus on the dark energy EoS parameter $w$. Building on methodologies outlined in \cite{Yan_2020}, the specific techniques are described as follows:

For a GW source propagating from direction \(\hat{\Omega}\), the induced timing residual at time \(t\) is expressed as:
\begin{equation}
	s(t, \hat{\Omega}) = F^{+}(\hat{\Omega}) \Delta A_{+}(t) + F^{\times}(\hat{\Omega}) \Delta A_{\times}(t),
	\label{eq:residual}
\end{equation}
where \(F^{+}(\hat{\Omega})\) and \(F^{\times}(\hat{\Omega})\) are antenna pattern functions given by:
\begin{align}
	F^{+}(\hat{\Omega}) &= \frac{1}{4(1 - \cos\theta)} \Bigl[ (1 + \sin^2\delta)\cos^2\delta_p \cos[2(\alpha - \alpha_p)] \nonumber \\
	&\quad - \sin2\delta \sin2\delta_p \cos(\alpha - \alpha_p) + \cos^2\delta(2 - 3\cos^2\delta_p) \Bigr], \\
	F^{\times}(\hat{\Omega}) &= \frac{1}{2(1 - \cos\theta)} \Bigl[ \cos\delta \sin2\delta_p \sin(\alpha - \alpha_p) \nonumber \\
	&\quad - \sin\delta \cos^2\delta_p \sin[2(\alpha - \alpha_p)] \Bigr].
\end{align}
Here, \((\alpha, \delta)\) and \((\alpha_p, \delta_p)\) denote the right ascension (RA) and declination (DEC) of the GW source and pulsar, respectively, and \(\theta\) is their angular separation:
\begin{equation}
	\cos\theta = \cos\delta \cos\delta_p \cos(\alpha - \alpha_p) + \sin\delta \sin\delta_p.
	\label{eq:angular_separation}
\end{equation}
The terms \(\Delta A_{[+,\times]}(t) = A_{[+,\times]}(t) - A_{[+,\times]}(t_p)\) incorporate the Earth term \(A_{[+,\times]}(t)\) and pulsar term \(A_{[+,\times]}(t_p)\), where \(t_p = t - d_p(1 - \cos\theta)/c\) is the time when the GW passes the MSP with a pulsar distance \(d_p\). For circular SMBBHs, these amplitudes are:
\begin{align}
	A_{+}(t) &= \frac{h_0(t)}{2\pi f(t)} \Bigl\{ (1 + \cos^2\iota) \cos2\psi \sin[\phi(t) + \phi_0] \nonumber \\
	&\quad + 2\cos\iota \sin2\psi \cos[\phi(t) + \phi_0] \Bigr\}, \\
	A_{\times}(t) &= \frac{h_0(t)}{2\pi f(t)} \Bigl\{ (1 + \cos^2\iota) \sin2\psi \sin[\phi(t) + \phi_0] \nonumber \\
	&\quad - 2\cos\iota \cos2\psi \cos[\phi(t) + \phi_0] \Bigr\}.
\end{align}
Here, \(\iota\) is the inclination angle, \(\psi\) the polarization angle, \(\phi_0\) the phase constant (these three angles are randomly assigned to our SMBBH dataset  according to certain distribution, see Table~\ref{Tab:angledis}) and \(h_0\) the GW strain amplitude given by:
\begin{equation}
	h_0 = \frac{2 \left( G \mathcal{M}_c^z \right)^{5/3}}{c^4} \frac{(\pi f)^{2/3}}{d_L},
	\label{eq:strain}
\end{equation}
where \(\mathcal{M}_c^z = \mathcal{M}_c(1+z)\) is the redshifted chirp mass, \(d_L\) the luminosity distance, and \(f\) the observed GW frequency. The rest-frame frequency \(f_r\) relates to \(f\) via \(f = f_r/(1+z)\),  and the GW frequency and phase evolve as:
\begin{align}
	f(t) &= \left[ f_0^{-8/3} - \frac{256}{5}\pi^{8/3} \left( \frac{G\mathcal{M}_c^z}{c^3} \right)^{5/3} t \right]^{-3/8}, \\
	\phi(t) &= \frac{1}{16} \left( \frac{G\mathcal{M}_c^z}{c^3} \right)^{-5/3} \left[ (\pi f_0)^{-5/3} - (\pi f(t))^{-5/3} \right].
\end{align}

The SNR for a PTA with \(N_p\) pulsars is computed as:
\begin{equation}
	\rho^2 = \sum_{j=1}^{N_p} \sum_{i=1}^{N} \left( \frac{s_j(t_i)}{\sigma_{i,j}} \right)^2,
	\label{eq:snr}
\end{equation}
where \(N\) is the number of data points per pulsar, \(s_j(t_i)\) the timing residual for the \(j\)-th pulsar, and \(\sigma_{i,j}\) the RMS timing noise. Parameter estimation employs the Fisher information matrix:
\begin{equation}
	\Gamma_{ab} = \sum_{j=1}^{N_p} \sum_{i=1}^{N} \frac{1}{\sigma_{i,j}^2} \frac{\partial s(t_i)}{\partial p_a} \frac{\partial s(t_i)}{\partial p_b},
	\label{eq:fisher}
\end{equation}
where \(p_a\) and \(p_b\) denote parameters (e.g., \(d_L\), \(\mathcal{M}_c^z\), \(\iota\)). \(P(\iota) \propto \sin\iota\) adds a prior as \(\Gamma_{\iota\iota} \to \Gamma_{\iota\iota} + 1/\sin^2\iota\). Parameter uncertainties can be derived from \(\Delta p_a = \sqrt{(\Gamma^{-1})_{aa}}\). If electromagnetic counterpart can be observed, further priors to the ith parameter can be incorporated via \(\Gamma_{ab} \to \Gamma_{ab} + \delta_{ab}/\sigma_i^2\).

The luminosity distance \(d_{\rm L}\) of a GW source in a flat universe is fundamentally linked to cosmological parameters through the relation:
\begin{equation}
	d_{\rm L} = (1+z) \int_0^z \frac{dz'}{H(z')},
\end{equation}
where \(H(z)\)—the Hubble parameter—encapsulates the dark energy EoS parameter \(w\) via:
\begin{equation}
	H(z) = H_0 \left[ \Omega_{\rm m}(1+z)^3 + \Omega_{\rm de}(1+z)^{3(1+w)} \right]^{1/2}.
	\label{eq:hubble_param}
\end{equation}
CMB observations provide tight constraints on (\(H_0\), \(\Omega_{\rm m}\), \(\Omega_{\rm de}\)). Following \cite{Yan_2020}, we treat these parameters as fixed by CMB data. For individual SMBBHs, the uncertainty in \(w\) is derived from luminosity distance measurements:
\begin{equation}
	\Delta w = d_{\rm L} \left| \frac{\partial d_{\rm L}}{\partial w} \right|^{-1} \frac{\sigma_{d_{\rm L}}}{d_{\rm L}},
	\label{eq:delta_w_single}
\end{equation}
where \(\sigma_{d_{\rm L}}\) incorporates both GW measurement errors (\(\Delta d_{\rm L}\)) and weak lensing uncertainties (\(\tilde{\Delta}d_{\rm L}\)) modeled as:
\begin{equation}
	\tilde{\Delta}d_{\rm L} = d_{\rm L} \times 0.066 \left( \frac{1 - (1+z)^{-0.25}}{0.25} \right)^{1.8}.
\end{equation}
For a population of \(N\) detected SMBBHs, the combined constraint on \(w\) is given by:
\begin{equation}
	\Delta w = \left[ \sum_{i=1}^{N} \left( \Delta w_i \right)^{-2} \right]^{-1/2},
	\label{eq:delta_w_combined}
\end{equation}
where \(\Delta w_i\) denotes the uncertainty from the \(i\)-th source.

It should be noted that in this work, we specifically consider a constant equation of state for dark energy, that is, the parameter $w$ is assumed to be a constant and does not evolve with redshift. We will briefly address how our methodology can be extended to investigate dynamical dark energy in the Discussion section (Sec. \ref{sec:concl}).

\section{Dataset}
\label{sec:dataset}
\subsection{Simulation Setup and Light-Cone Construction}
We employ the \texttt{L-Galaxies} semi-analytic galaxy formation model Guo2013a \cite{2011MNRAS.413..101G,2013MNRAS.428.1351G} implemented on the Millennium simulation database \cite{MillenniumDB} to reconstruct galaxy assembly histories within a comoving volume of $500\,\mathrm{Mpc}/h_0$ spanning $z > 14$ to $z = 0$. We construct light-cone catalogs with methods similar to that of \cite{Yang:2024mqz}, i.e., we apply periodic boundary conditions to replicate the simulation box, generating an effectively infinite cosmic volume. Then for each merger-formed galaxy recorded at certain redshift snapshot in each simulation box, galaxy merger times are randomly assigned between the current redshift snapshot and the previous one. We assume SMBBH formation is tied to galaxy mergers, with the time of entry into the PTA frequency band ($10^{-9} - 10^{-6}\,\mathrm{Hz}$) determined by:
\begin{equation}
	t_{\rm entry} = t_{\rm galaxy} + \tau_H,
	\label{eq:t_entry}
\end{equation}
where $t_{\rm galaxy}$ is the galaxy merger time, and $\tau_H$ represents the post-merger hardening timescale for the SMBBH. We investigate three hardening scenarios: $\tau_H = \{0.1,\ 5,\ 10\}\,\mathrm{Gyr}$.
The coalescence time of the SMBBH is determined by:
\begin{equation}
	t_{\rm c} = t_{\rm entry} + \tau_{\rm evo},
	\label{eq:t_coa}
\end{equation}
with the remaining evolution time to coalescence $\tau_{\rm evo}$ for the SMBBH calculated via:
\begin{equation}
	\tau_{\rm evo} = \frac{5}{256} (\pi f_r)^{-8/3} \left( \frac{G M_c}{c^3} \right)^{-5/3}
	\label{eq:timetocoa}
\end{equation}
here $M_c \equiv (m_1 m_2)^{3/5}/(m_1 + m_2)^{1/5}$ is the chirp mass, and $f_r=(1+z)10^{-9}\,\mathrm{Hz}$ is the rest-frame frequency. Suppose $z_{\rm entry}$ and $z_c$ are the redshifts related to $t_{\rm entry}$ and $t_c$ respectively, events whose distance to Earth  satisfying:
\begin{equation}
	d_c(z_{\rm c}) \leq d_c \leq d_c(z_{\rm entry}) 
\end{equation}
are retained as potential observable light-cone sources ($d_c(z)$  denotes the comoving distance at redshift $z$), since the look-back time determined by this distance allows Earth-based observers to detect this GW event during its evolutionary phase where it has entered the PTA frequency band but has not yet undergone coalescence. By calculating how much time remains from this look-back time to the time of coalescence, the rest frame frequency of each binary black hole event can be determined by solving for $f_r$ with a certain $\tau_{\rm evo}$ through Equation~\eqref{eq:timetocoa}. Additionally, redshift of the SMBBH can be derived from its distance to Earth, and the GW strain amplitudes can be calculated using Equation \eqref{eq:strain}.

\begin{table}
\caption{The distribution of random angular parameters.
}

\begin{center}
	\begin{tabularx}{1\columnwidth }{>{\centering\arraybackslash}X >{\centering\arraybackslash}X >{\centering\arraybackslash}X}
		\hline \hline
		Parameter & Range & Distribution  \\\hline
		Polarization angle & $\psi \in [0,\pi]$ rad & Uniform \\ 
		Initial phase & $\phi_0 \in [0,2\pi]$ rad  &  Uniform \\
		Inclination angule & $\iota \in [-\frac{\pi}{2},\frac{\pi}{2}]$ rad \hspace{0.5cm}& Sine \\ \hline \hline
	\end{tabularx}
	\label{Tab:angledis}
\end{center}
\end{table}

\begin{figure}
\centering 
\includegraphics[width=0.45\textwidth]{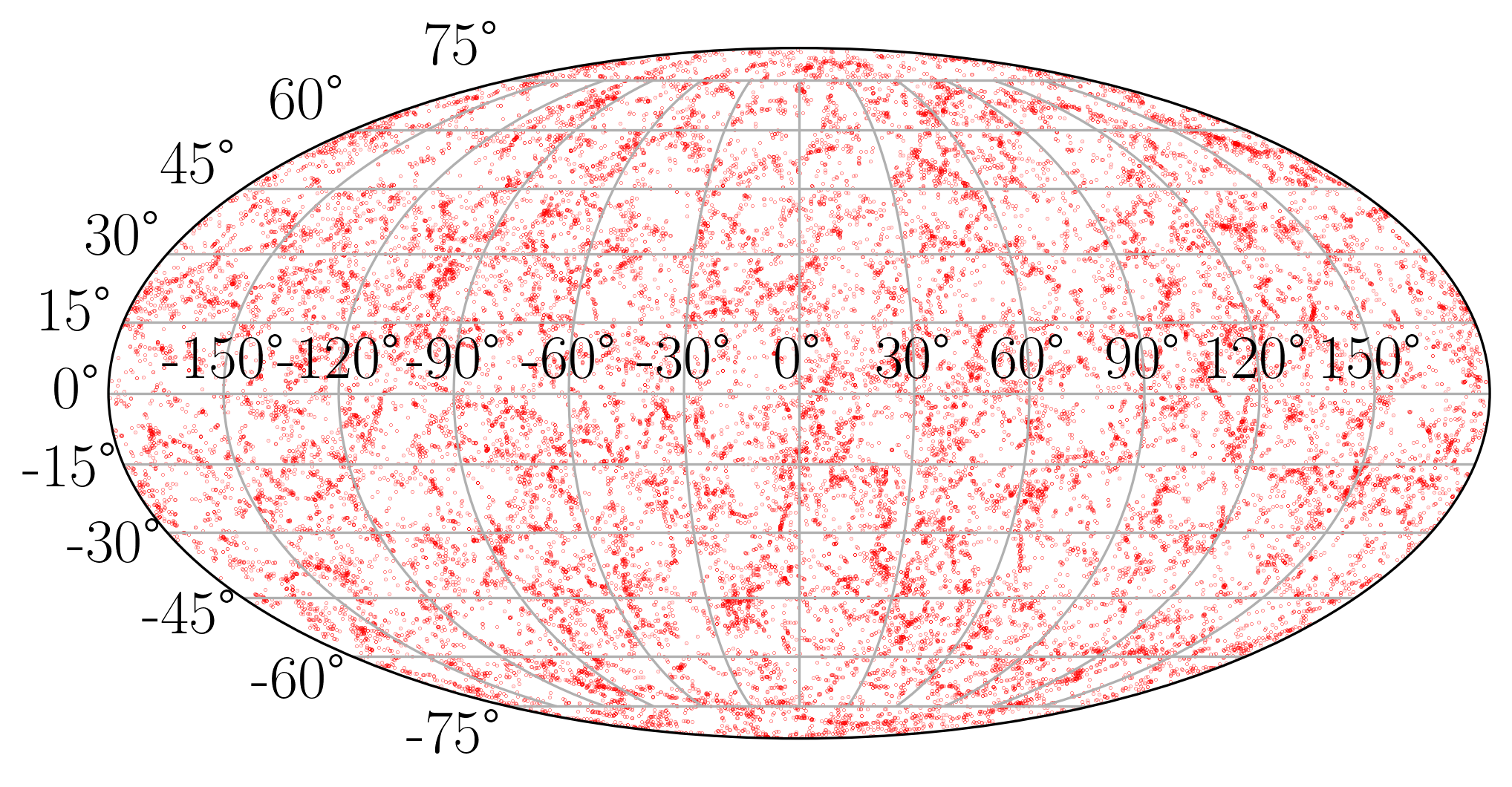}
\caption{The all-sky map of angular distribution for our light-cone GW sources around redshift $z=0.09$ in the $\tau_H=0.1$Gyr case.}
\label{fig:lightcone}
\end{figure}

For each $\tau_H$, we generate 50 realizations by randomizing Earth's position among host galaxies. The basic properties of the resulted light-cone populations are presented in Table~\ref{Tab:lightcone}. In Fig.~\ref{fig:lightcone},  we also present the angular distribution sky map of light-cone GW sources around redshift $z=0.09$ for the $\tau_H=0.1$ case to provide a more intuitive visualization of our light-cone events.

\begin{table*}
\caption{Fundamental properties of light-cone events across all three hardening timescales. Here $n_s$ denotes the total number of GW  sources in the light-cone population. $z_{\rm peak}$ and $z_{\rm max}$ represent the redshift of peak GW event density and the maximum observed redshift, respectively. Similarly, $M_{\rm c,peak}$ indicates the chirp mass at peak event density, while $M_c$ show the characteristic distribution range of chirp masses. Finally, $h_c$ denotes the average characteristic strain amplitude from the superposition of GW signals at frequency  $\rm yr^{-1}$.
}

\begin{center}
\begin{tabularx}{0.8\textwidth}{ 
		c|  
		*{6}{>{\centering\arraybackslash}X} 
	} 
	\hline \hline
	$\tau_{H}$ & $n_s$ & $z_{\text{peak}}$ & $z_{\text{max}}$ & $M_{c,\text{peak}}/M_{\odot}$ & $M_c/M_{\odot}$ & $h_c$ \\ 
	\hline
	0.1 Gyr    & $\sim 1 \times 10^{8}$  & $\sim 0.7$  & 6.2  & $\sim 10^7$          & $10^6{-}10^8$   & $4.10 \times 10^{-16}$ \\ 
	5 Gyr      & $\sim 6 \times 10^6$    & $\sim 0.3$  & 1.1  & $\sim 10^7$          & $10^5{-}10^9$   & $3.26 \times 10^{-16}$ \\
	10 Gyr     & $\sim 1 \times 10^4$    & $\sim 0.06$ & 0.2  & $\sim 10^7$          & $10^4{-}10^9$   & $2.22 \times 10^{-18}$ \\ 
	\hline \hline
\end{tabularx}
\label{Tab:lightcone}
\end{center}
\end{table*}

\subsection{PTA Sensitivity Analysis}
We compute SNR for next-generation PTAs using Equation~\eqref{eq:snr} for all the light-cone events from all realizations assuming a 14-day observing cadence. We select those bright GW sources with SNR$>8$ from our light-cone dataset. Following \cite{2012PhRvD..86l4028B}, we also impose an upper limit of $\sim 2N_p/7$ on the number of individual sources to each frequency bin (bin width determined by the inverse of observation duration) to determine the final resolvable sources. Our analysis explores parameter spaces of:
\begin{itemize}
\item Observational duration: $T_{\rm obs} \in \{10,\ 20,\ 30\}\,\mathrm{yr}$
\item Pulsar numbers: $N_p \in \{500,\ 1000\}$
\item Timing noise: $\sigma_t \in \{50,\ 100,\ 200\}\,\mathrm{ns}$
\end{itemize}
\begin{figure}
	\centering 
	\includegraphics[width=0.52\textwidth]{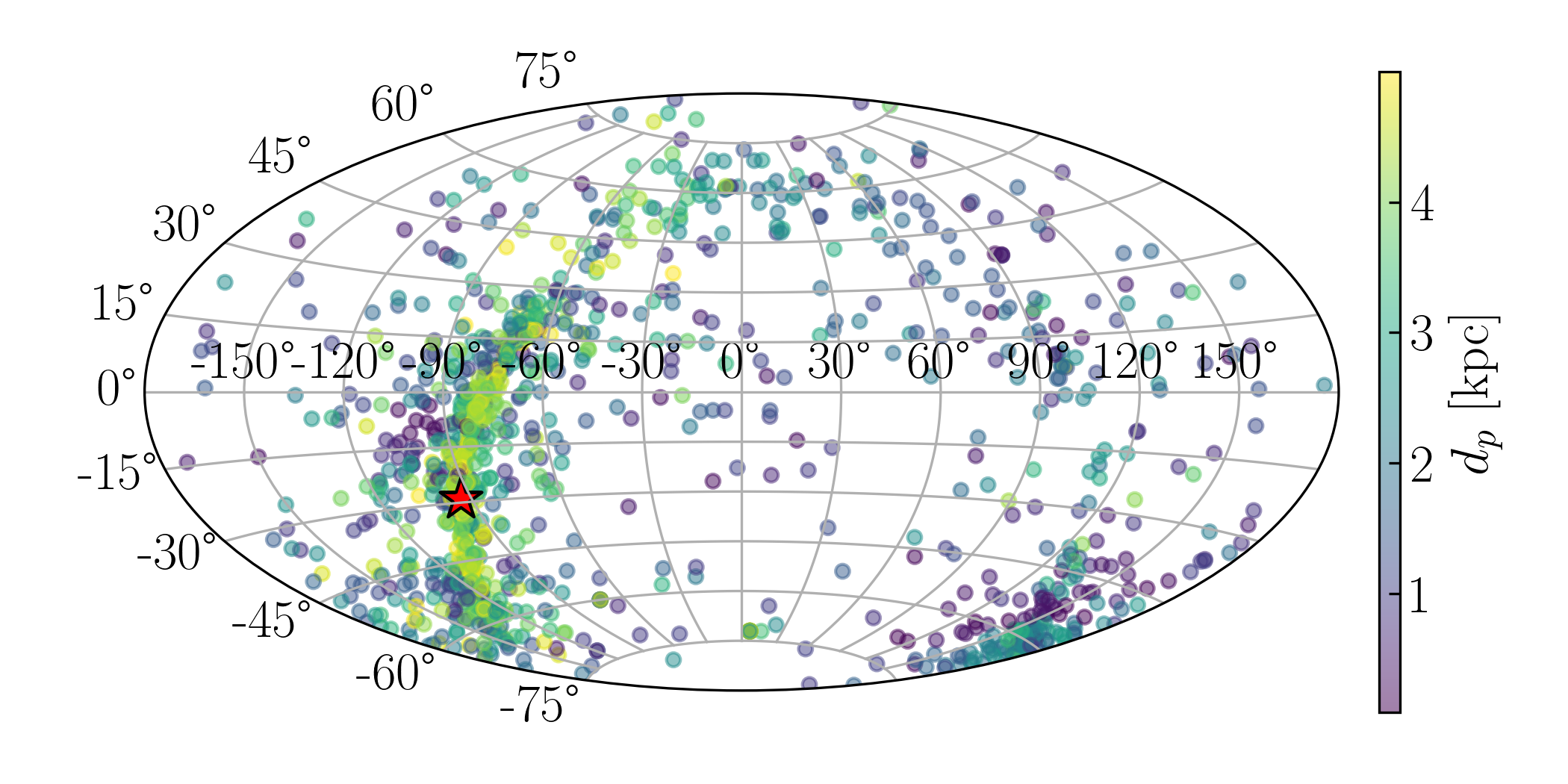}
	\caption{The all-sky map of angular distribution for MSPs, where colors are used to indicate the distances from these pulsars to Earth, and the red star symbol marks the direction of the Galactic Center.}
	\label{fig:pulsars}
\end{figure}

The pulsars we use are presented in Fig.~\ref{fig:pulsars}. They are generated according to ATNF pulsar catalog \footnote{Website: \url{http://www.atnf.csiro.au/people/pulsar/psrcat} \cite{2005AJ....129.1993M}}.

We conduct a systematic analysis of the number of resolvable individual GW sources across different realizations for all possible PTA parameter configurations in the above parameter space, and quantify the measurement precision achievable for constraining the dark energy EoS parameter $w$. We consider two scenarios. In the first scenario, all resolvable individual sources are assumed to have detectable electromagnetic counterparts, allowing precise determination of RA/DEC of these events. The masses of the SMBBH can be inferred from the masses of the host galaxies and incorporated as a prior into the Fisher information matrix. In the second scenario, only 10\% of the resolvable sources are assumed to have detectable electromagnetic counterparts. We randomly select 10\% of the SMBBH events in our sample to simulate events with detectable electromagnetic counterparts, and these selected ones are treated identically to the first scenario, while the remaining sources rely solely on pulsar timing residuals to estimate the detection accuracy of their parameters, without incorporating any host galaxy prior. For each realization, we perform 10 random selections for each of the 50 realization, thus a total of 500 results could be derived for each PTA parameter configuration and hardening time scale. The full numerical results are presented in Section~\ref{sec:results}.

\section{Results}
\label{sec:results}
\subsection{Resolvable Individual Sources}
In this subsection, we present the results of individual sources derived from different realizations of our methodology. As mentioned at the end of the previous section, we calculated the SNR for each selected light-cone events for all the realizations using Equation~\eqref{eq:snr} under various PTA parameter  configurations (including different numbers of pulsars, observation duration, and timing errors, totaling 18 combinations). We retained the bright individual events with SNR exceeding 8 under all these conditions and further analyzed the properties of their distribution.

\begin{figure*}
	\centering
	\includegraphics[width=0.32\textwidth]{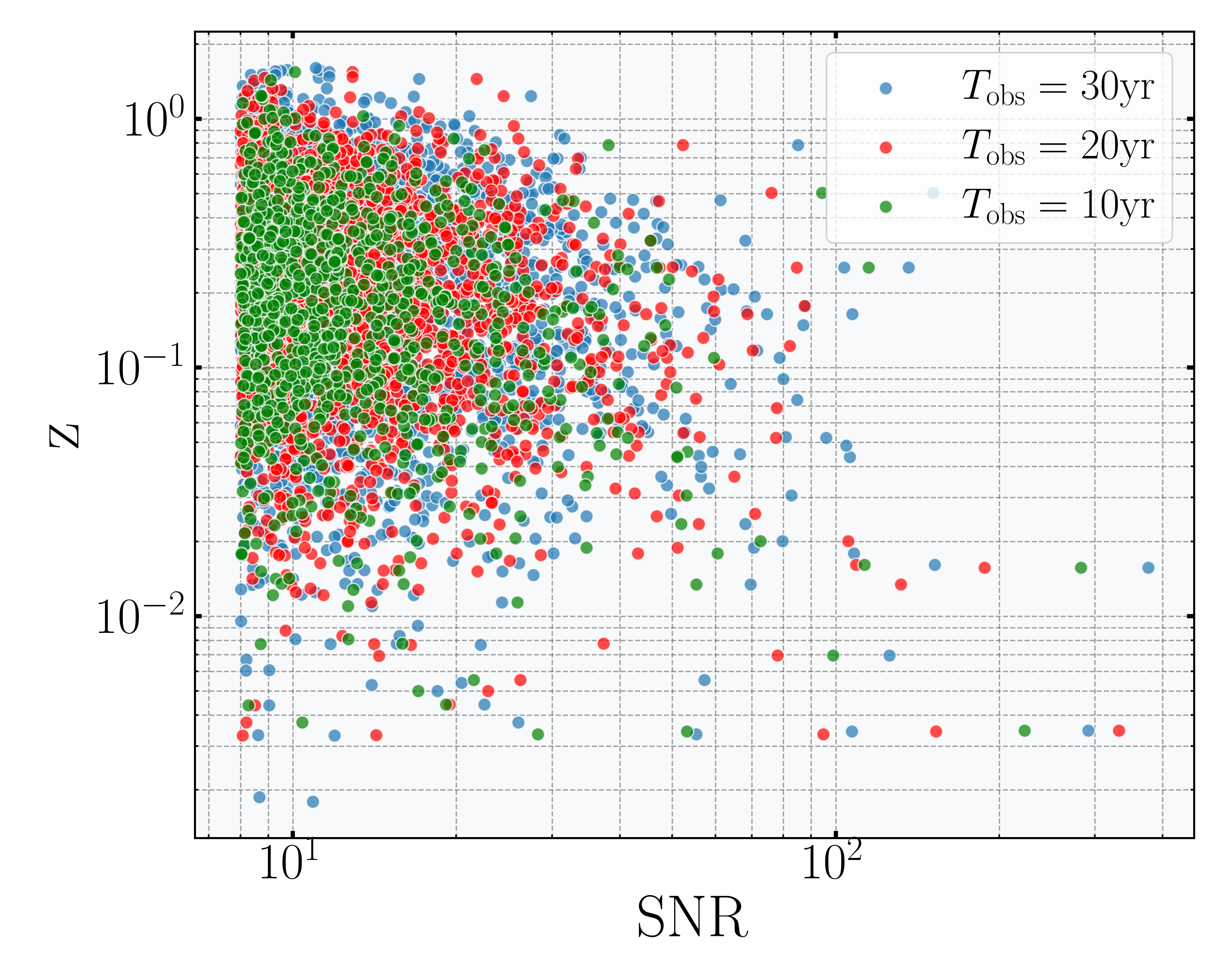}
	\hspace{-0.0cm}
	\includegraphics[width=0.32\textwidth]{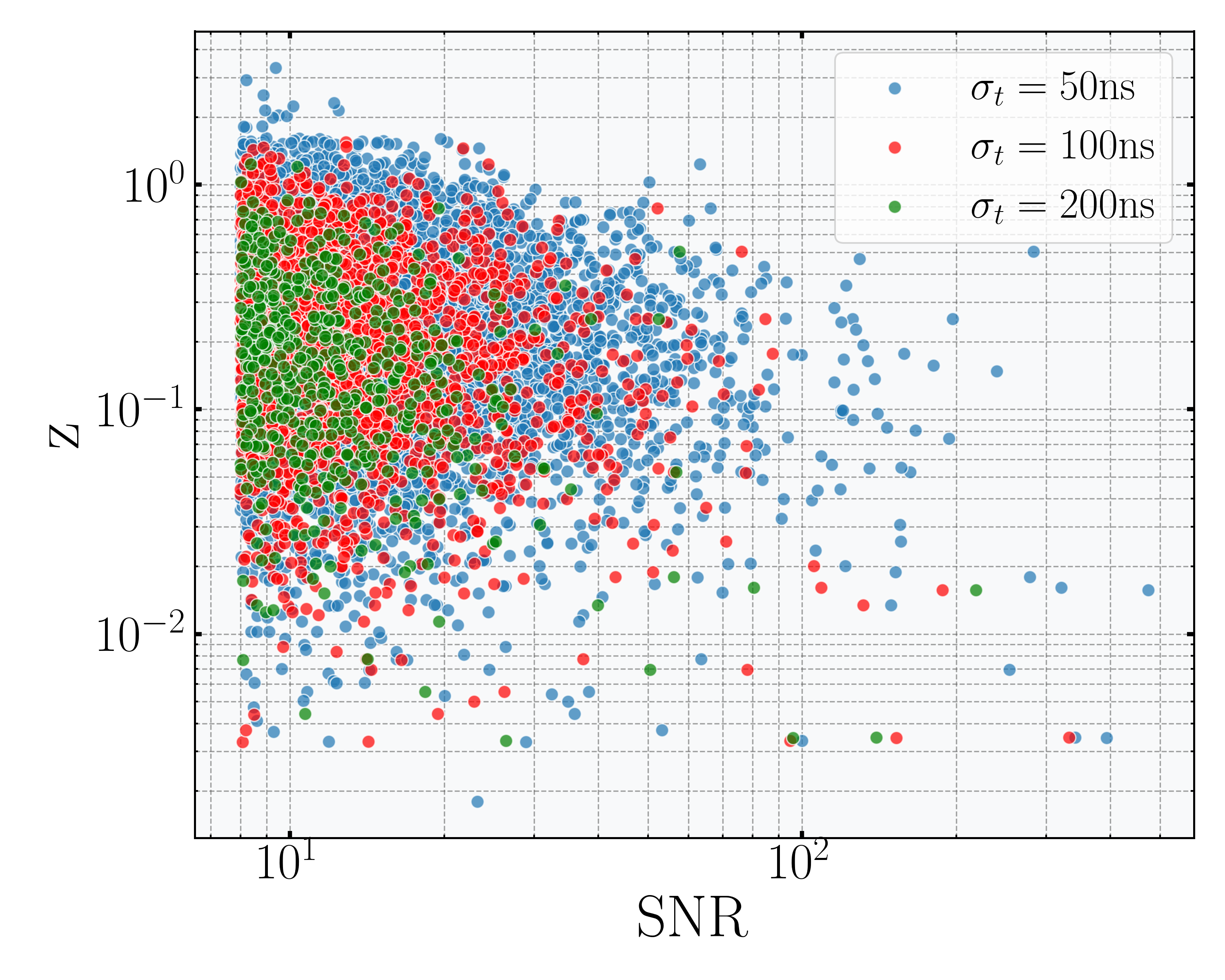}
	\hspace{-0.0cm}
	\includegraphics[width=0.32\textwidth]{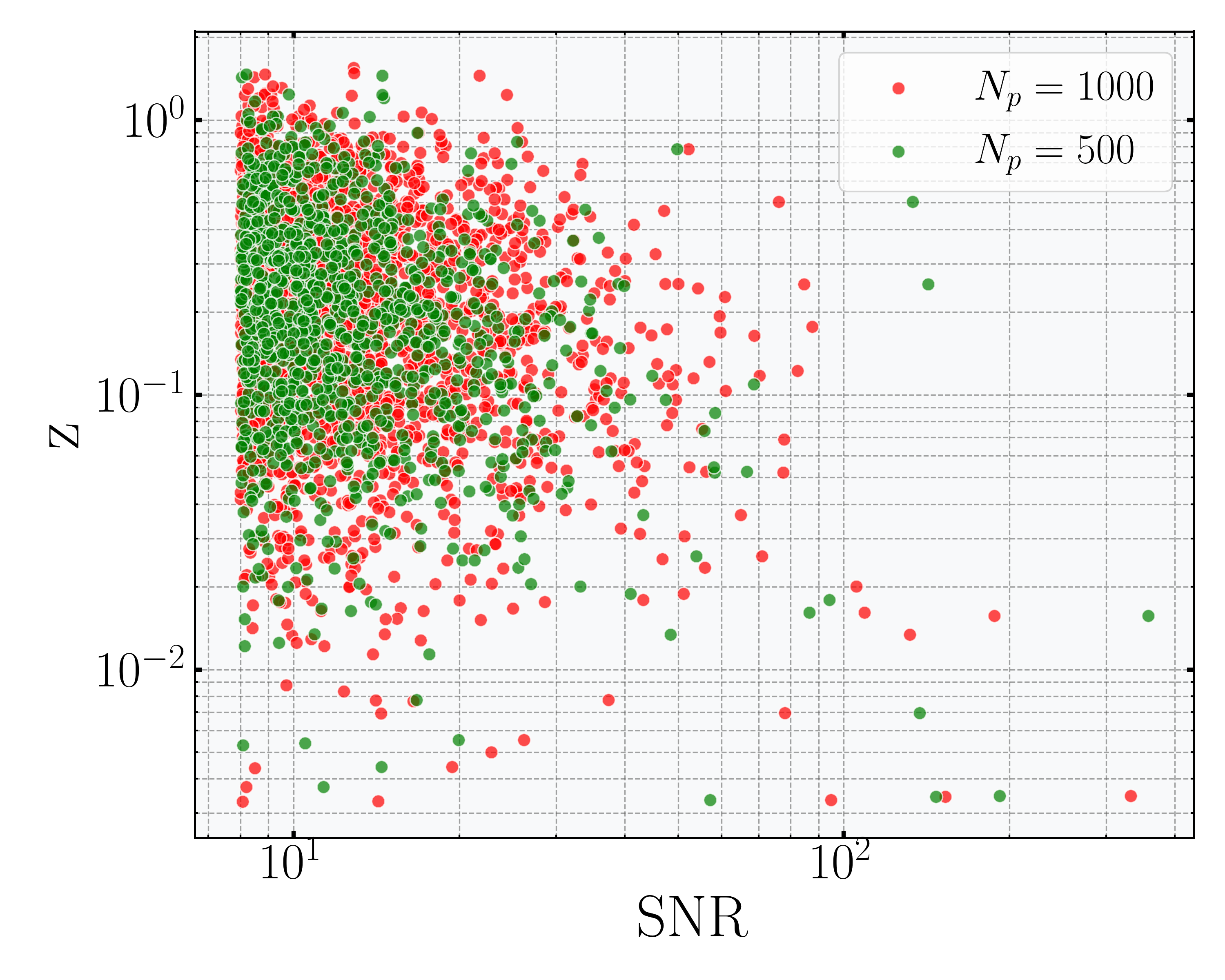}
	\hspace{-0.0cm}
	\includegraphics[width=0.32\textwidth]{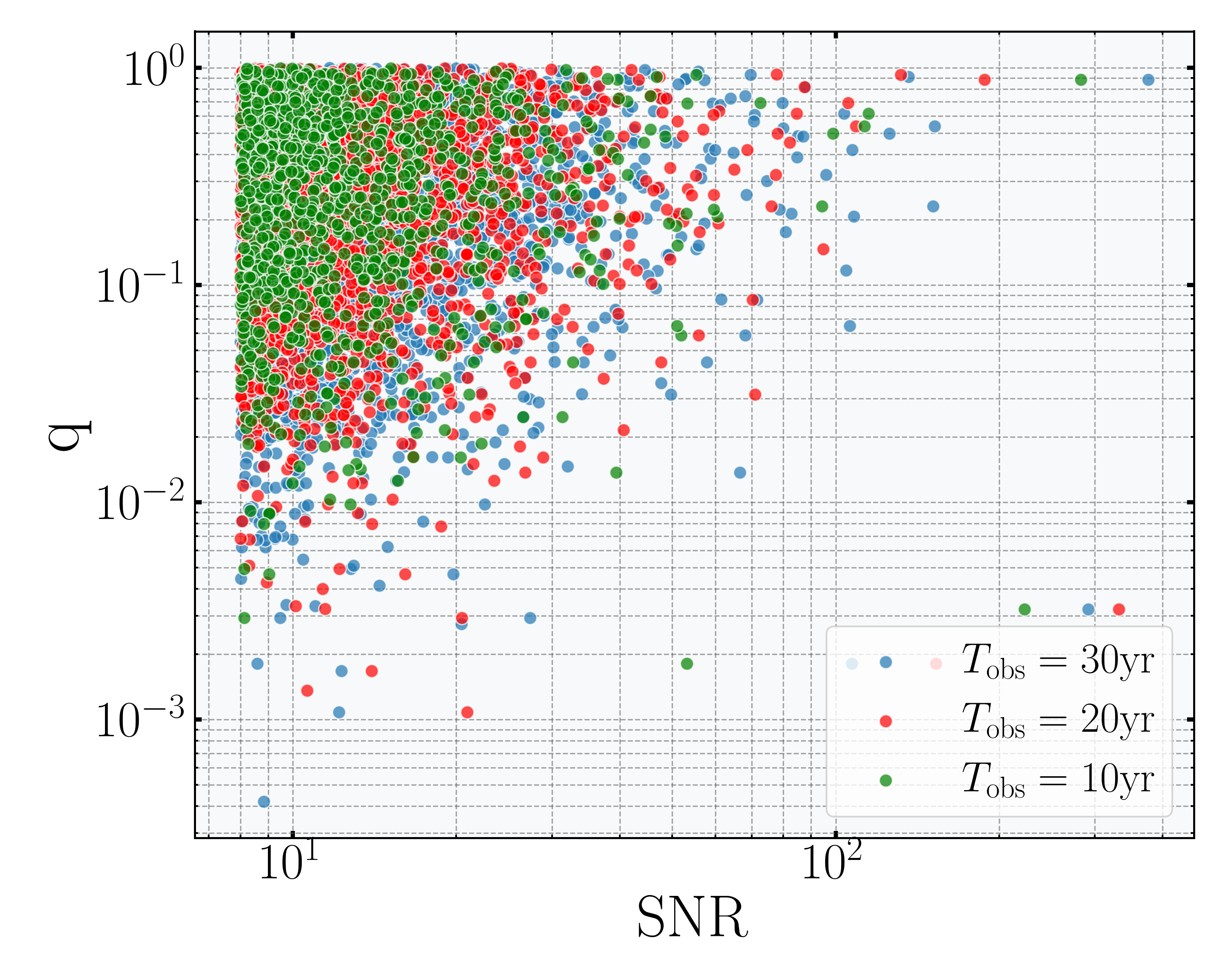}
	\hspace{-0.0cm}
	\includegraphics[width=0.32\textwidth]{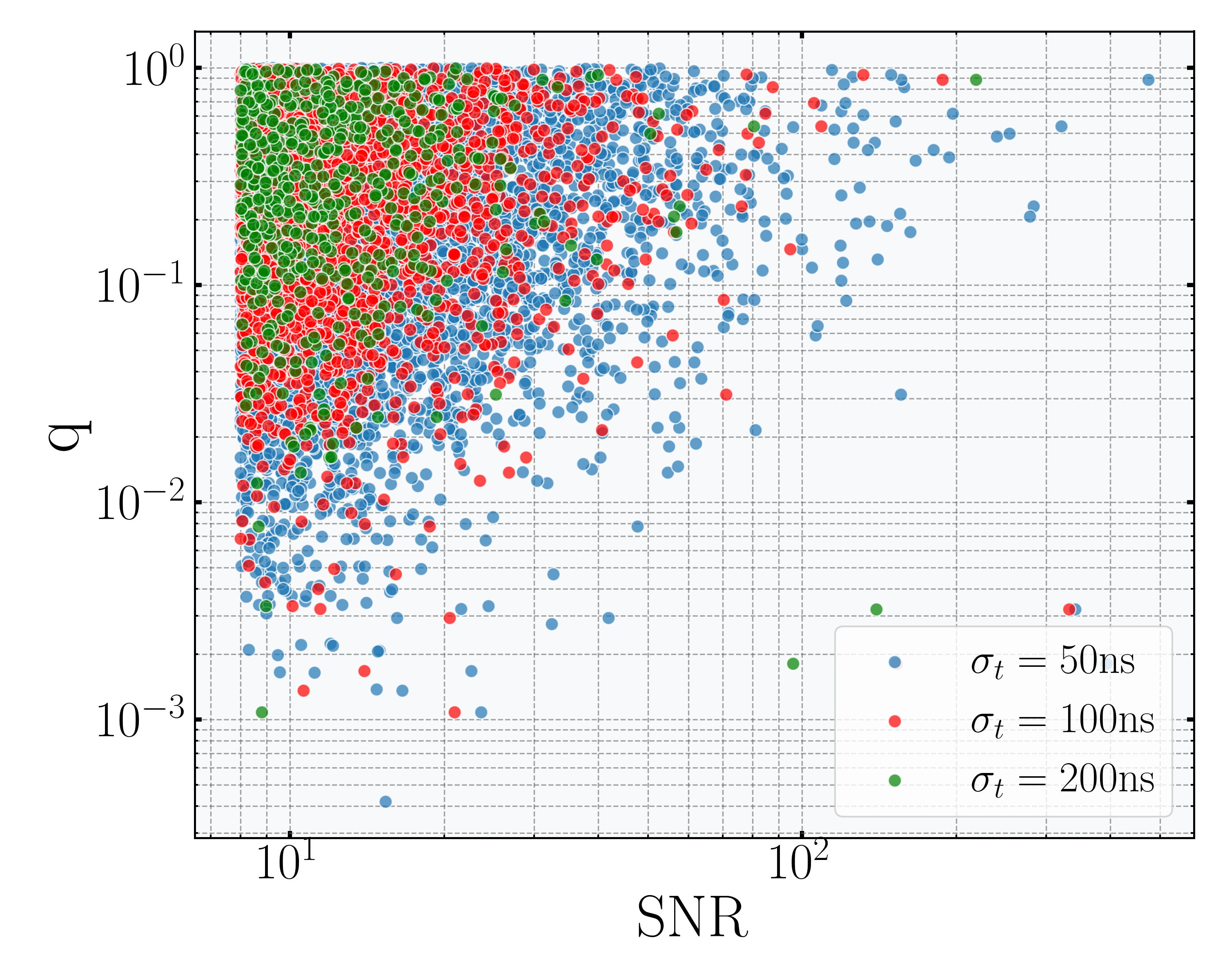}
	\hspace{-0.0cm}
	\includegraphics[width=0.32\textwidth]{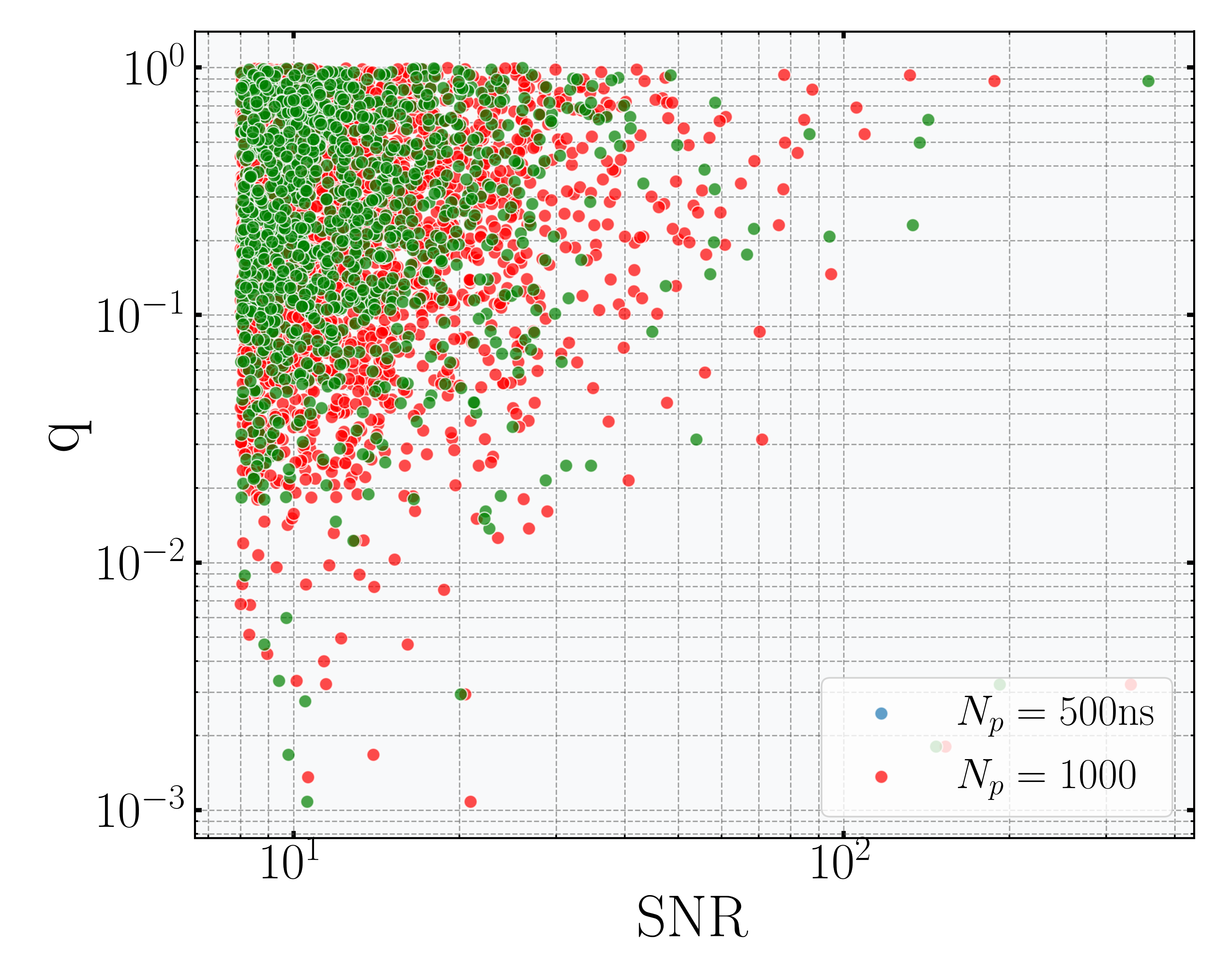}
	\hspace{-0.0cm}
	\includegraphics[width=0.32\textwidth]{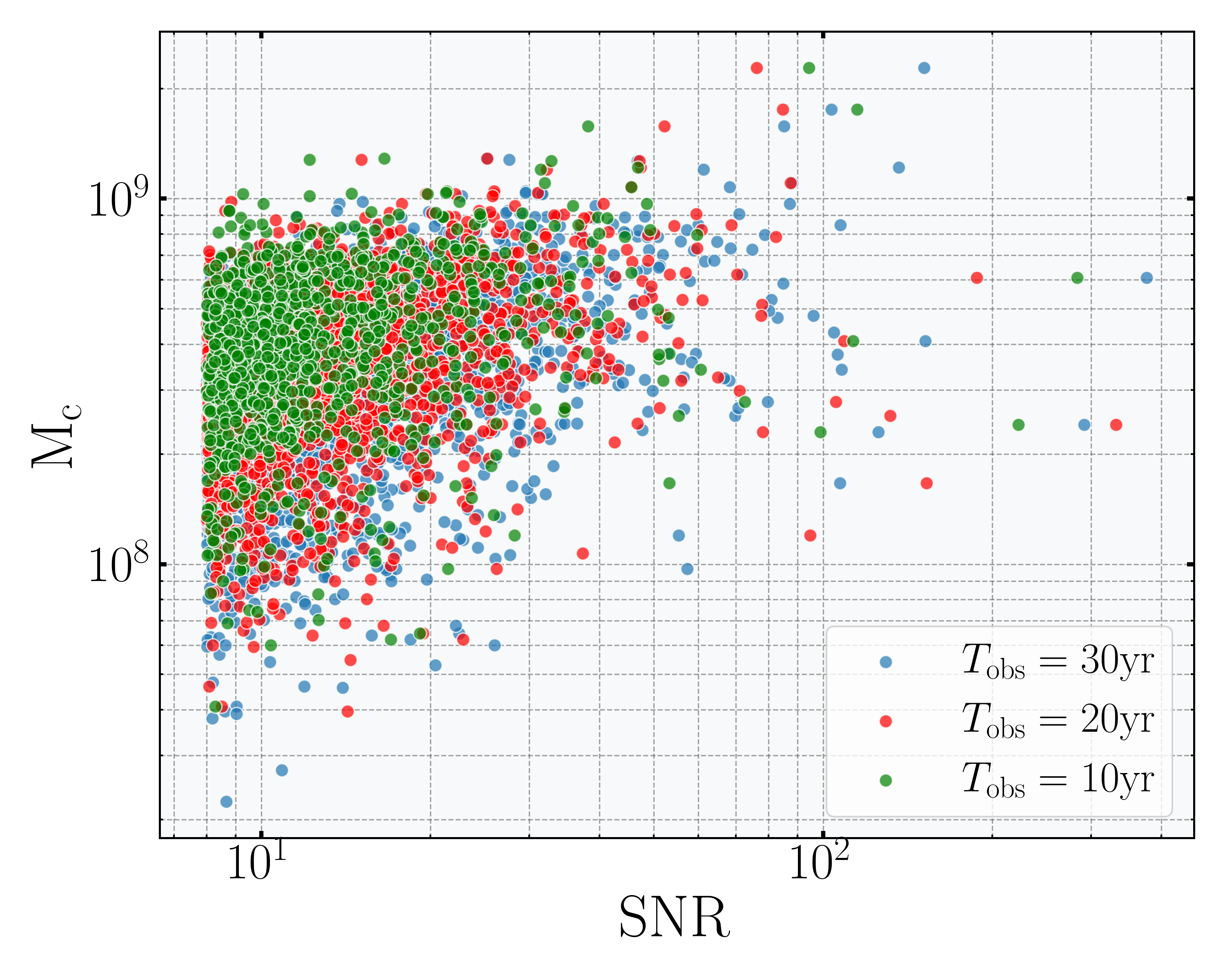}
	\hspace{-0.0cm}
	\includegraphics[width=0.32\textwidth]{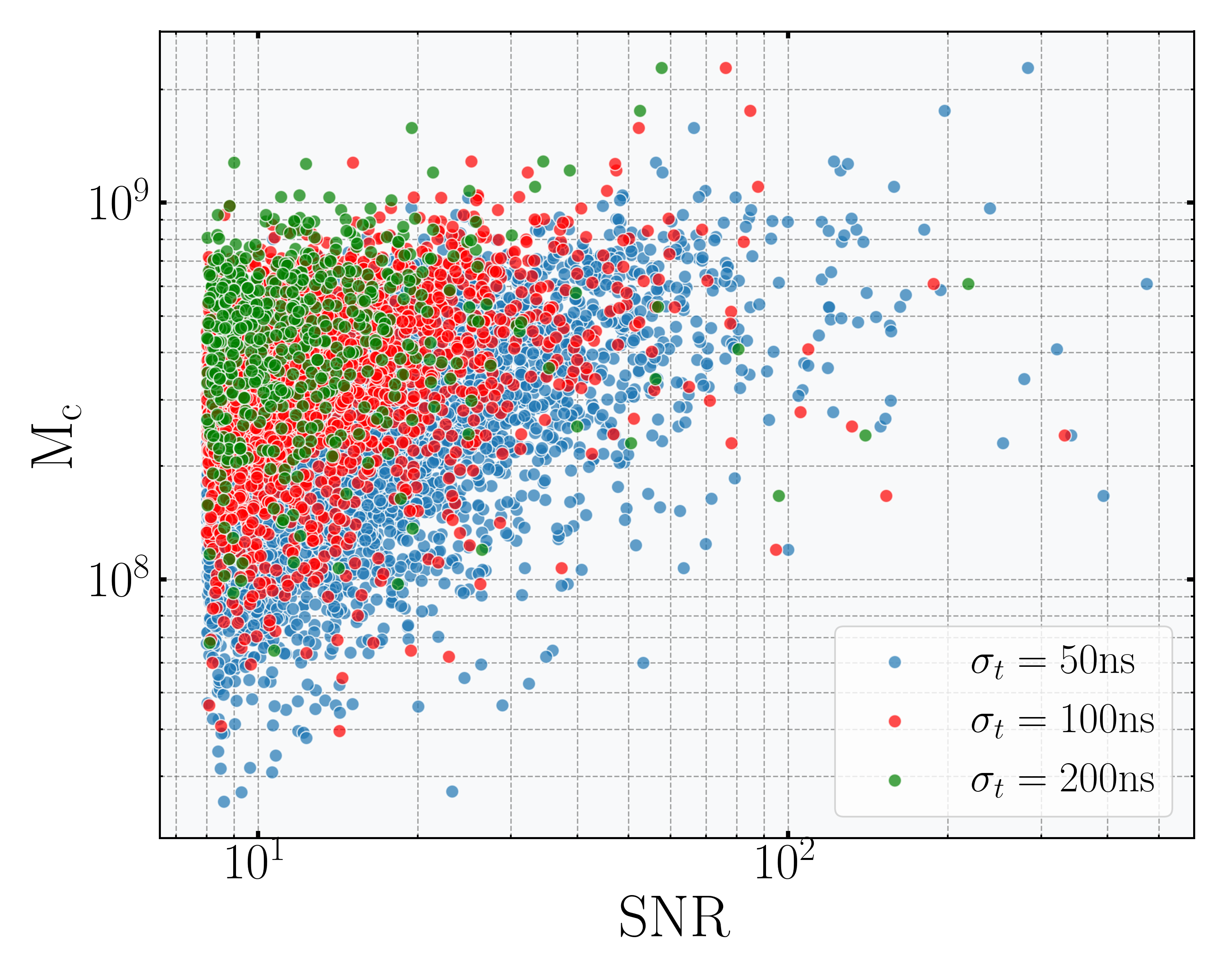}
	\hspace{-0.0cm}
	\includegraphics[width=0.32\textwidth]{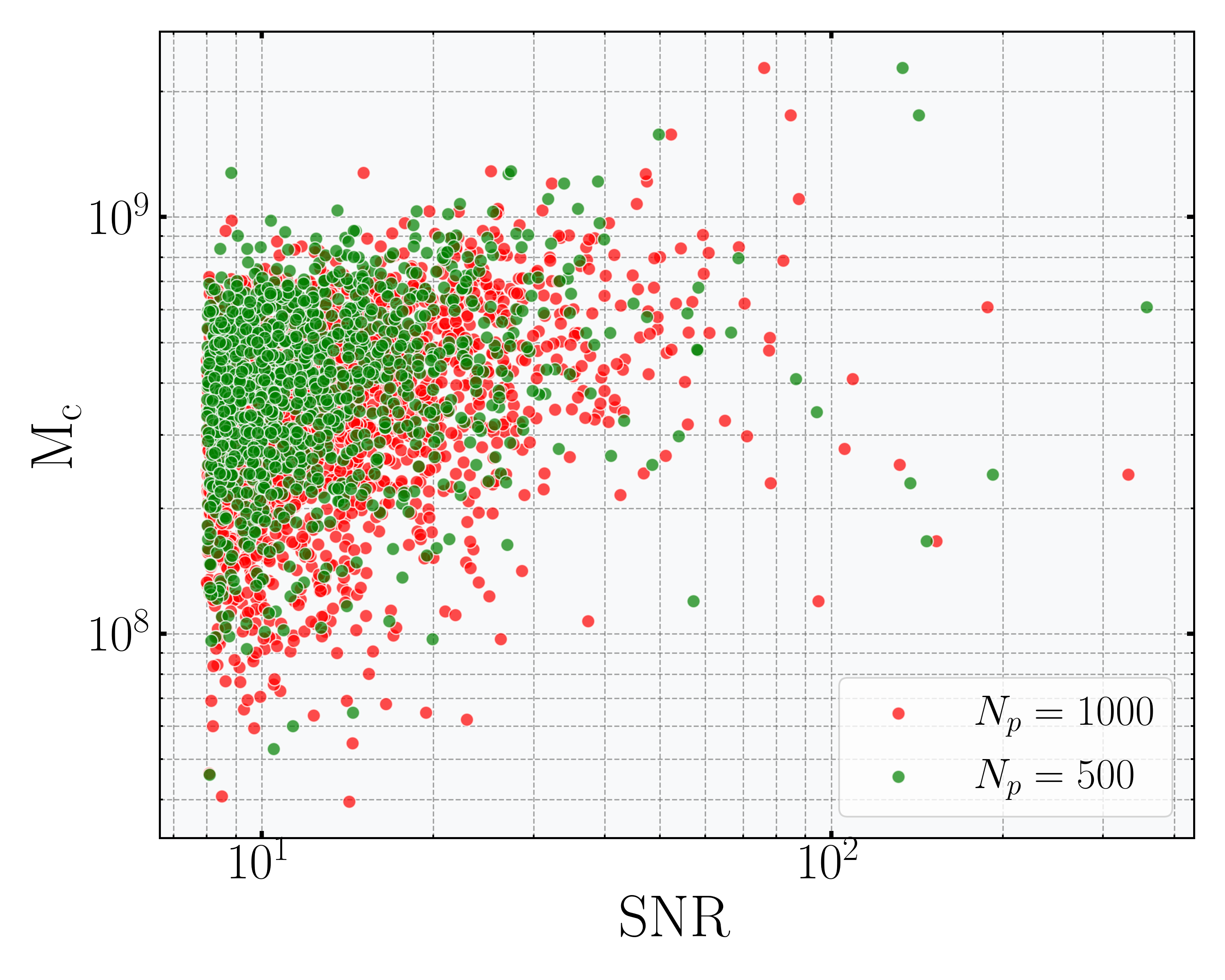}
	\caption{Scatter plots showing the distributions of SNR  versus redshift (top row), mass ratio of binary black holes (middle row), and chirp mass (bottom row) for all bright sources with SNR $>$ 8 in the most optimal case with hardening timescale $\tau_H=0.1\rm Gyr$. The left column displays results for different total observation duration (with other parameters set as $N_p$=1000, $\sigma_t$=100 ns); The middle column displays results under different timing error level (with other parameters set as $N_p=1000$, $T_{\rm obs}=20\rm yr$); The right column displays results for different numbers of pulsars (with other parameters set as $\sigma_t$=100 ns, $T_{\rm obs}=20\rm yr$).}
	\label{fig:scatter}
\end{figure*}

In Fig.~\ref{fig:scatter}, we present scatter plots showing the distribution of SNR versus redshift/mass ratio/chirp mass for individual sources from the most optimal case with $\tau_H=0.1\rm Gyr$ under different detector parameters. As shown in the figure, although our light-cone events can reach a maximum redshift of 6.2, the vast majority of individual sources with SNR exceeding 8 exhibit redshifts below 2. Meanwhile, the chirp masses of light cone events spans from $10^5M_\odot$ to $10^{10}M_\odot$, while the vast majority of bright individual sources exhibit chirp masses distributed between $10^8M_\odot$ to $10^{10}M_\odot$. As for mass ratio, the majority of events exhibit mass ratios distributed  between $10^{-2}$ and 1. It can also be seen from the figure that, events with higher SNR tend to exhibit larger chirp masses and mass ratios. However, regarding redshift, the number of bright sources peaks around $z=0.1$. This is understandable,  although low-redshift sources tend to be brighter, the relatively smaller comoving volume corresponds to lower redshifts leads to a corresponding reduction in the number of individual sources. As the observation time, timing accuracy, and number of pulsars improve, the number of bright individual sources increases, with timing accuracy appearing to have the most significant impact. 

\begin{figure*}
	\centering
	\includegraphics[width=0.32\textwidth]{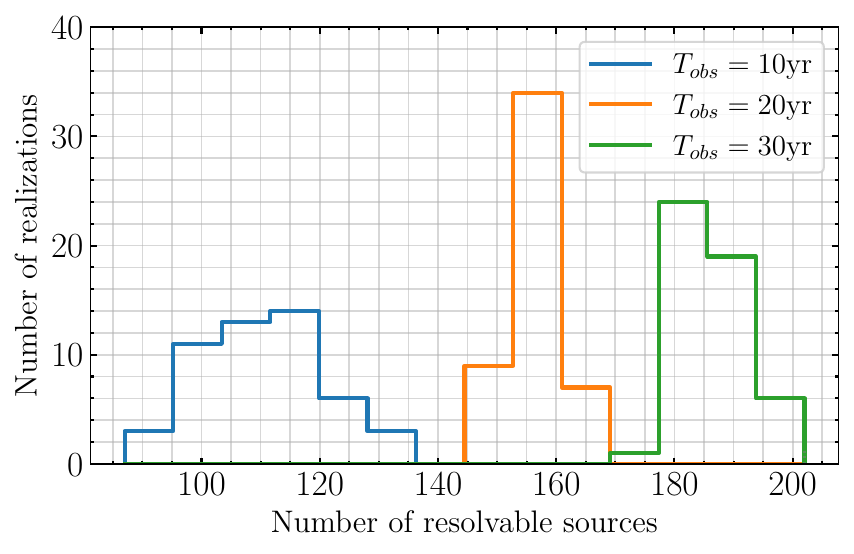}
	\hspace{-0.0cm}
	\includegraphics[width=0.32\textwidth]{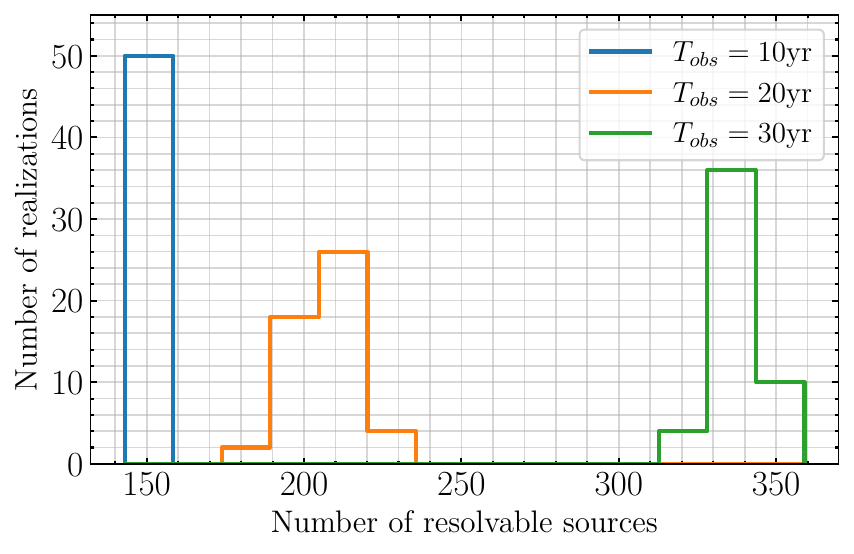}
	\hspace{-0.0cm}
	\includegraphics[width=0.32\textwidth]{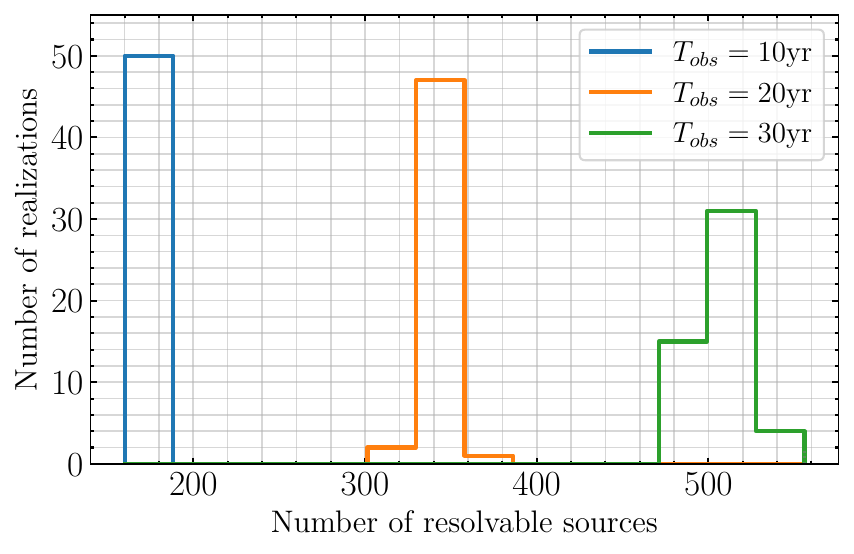}
	\hspace{-0.0cm}
	\includegraphics[width=0.32\textwidth]{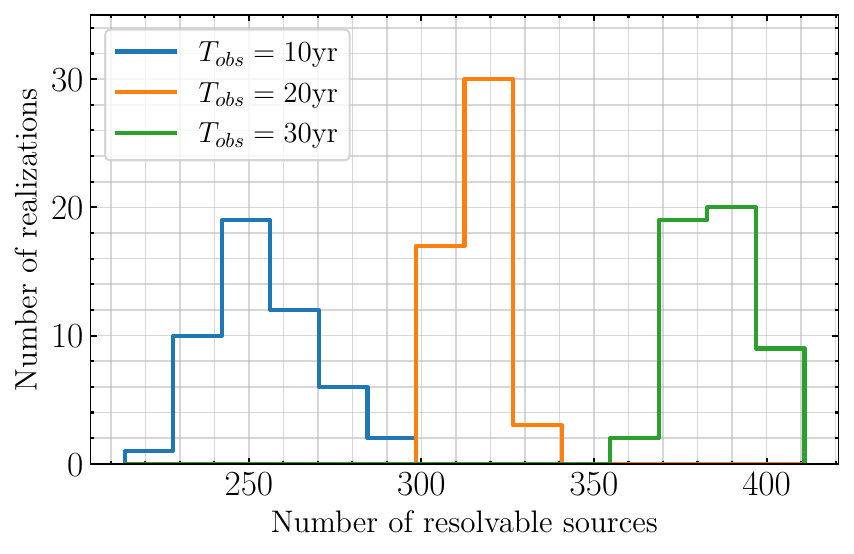}
	\hspace{-0.0cm}
	\includegraphics[width=0.32\textwidth]{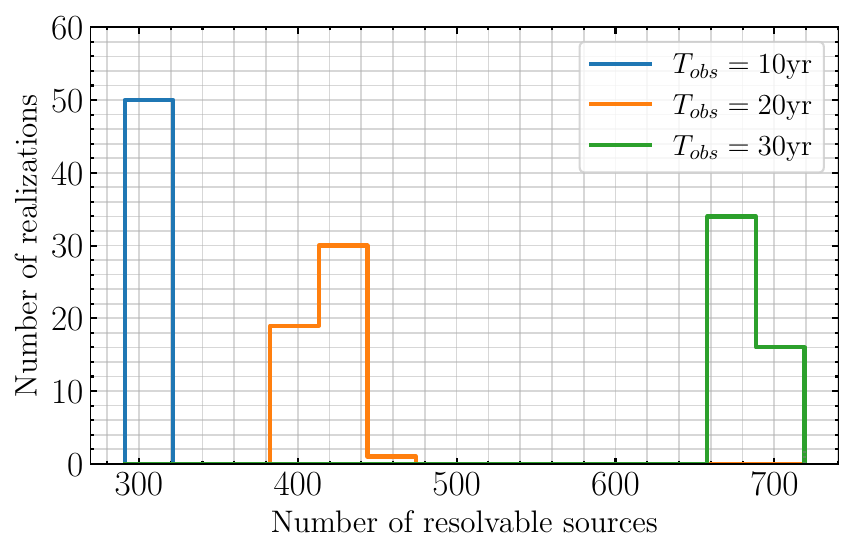}
	\hspace{-0.0cm}
	\includegraphics[width=0.32\textwidth]{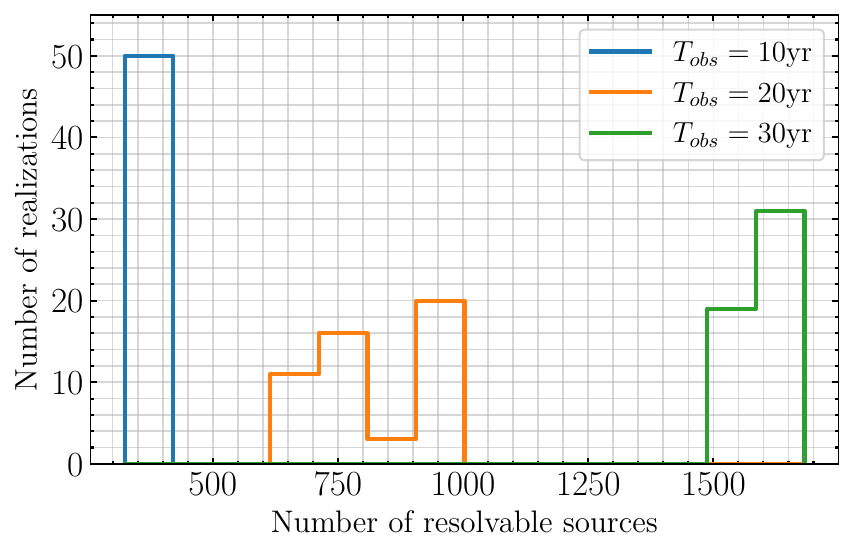}
	\caption{The number of realizations versus the number count of resolvable GW sources for different PTA parameter configurations in the $\tau_{H}=0.1$ Gyr case. The upper and lower panels represent the results for $N_p=500$ and $1000$, respectively, while the left, middle, and right panels correspond to results for $\sigma_t=200\rm ns$, $100\rm ns$ and $50\rm ns$ respectively.
	}
	\label{fig:numdistri1}
\end{figure*}

\begin{figure*}
	\centering
	\includegraphics[width=0.32\textwidth]{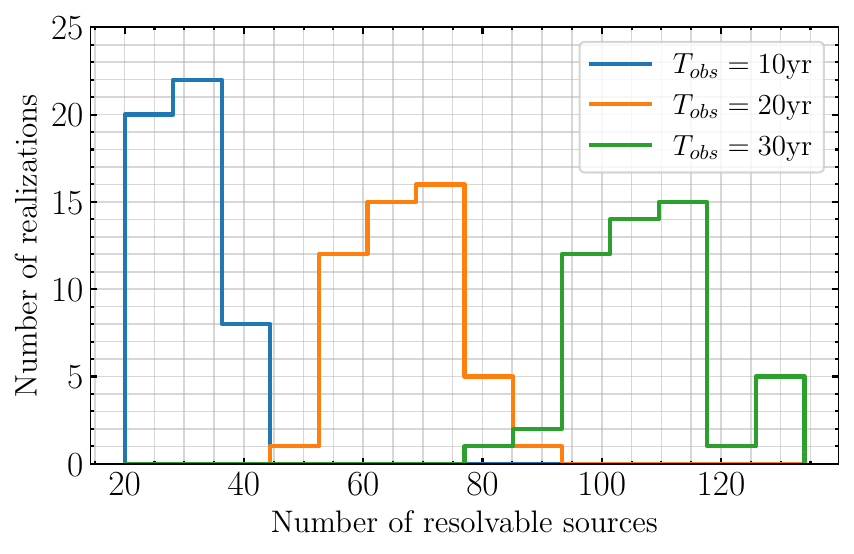}
	\hspace{-0.0cm}
	\includegraphics[width=0.32\textwidth]{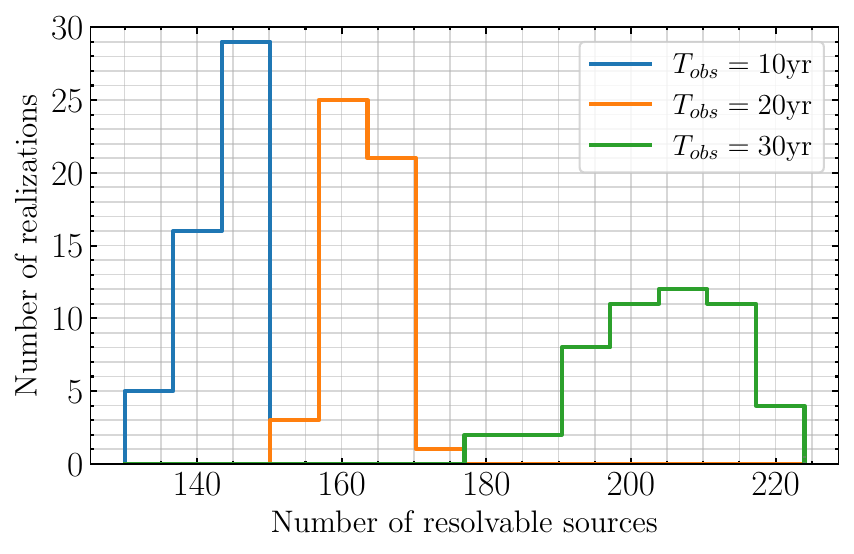}
	\hspace{-0.0cm}
	\includegraphics[width=0.32\textwidth]{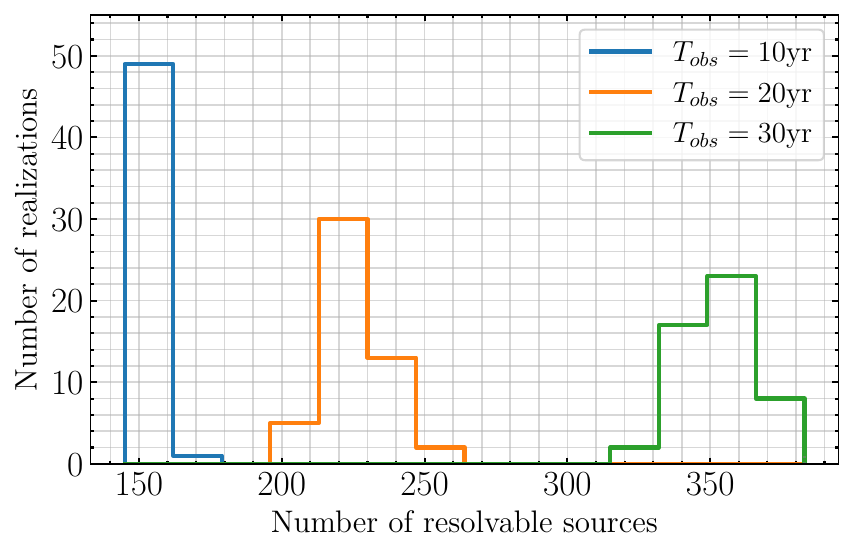}
	\hspace{-0.0cm}
	\includegraphics[width=0.32\textwidth]{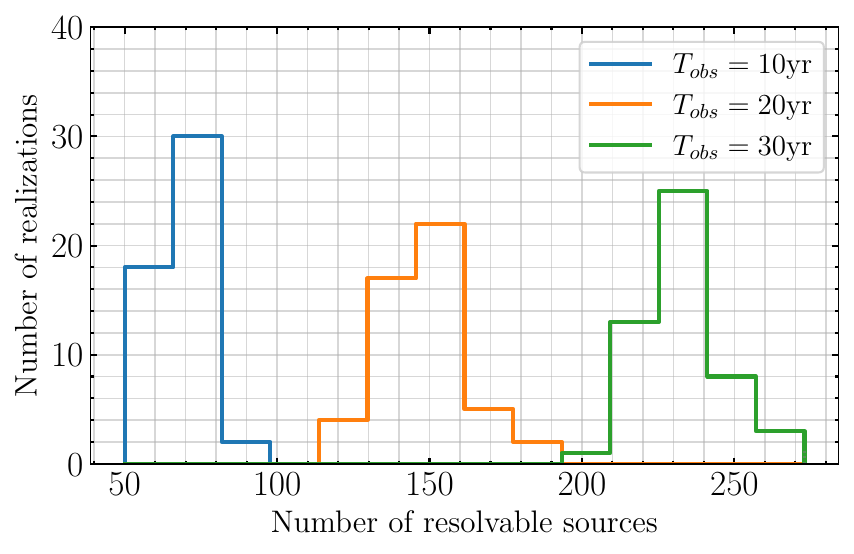}
	\hspace{-0.0cm}
	\includegraphics[width=0.32\textwidth]{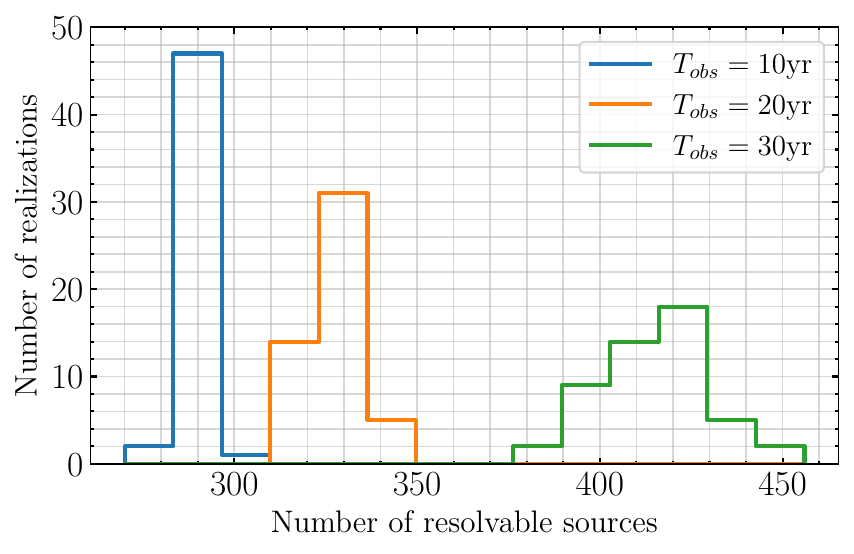}
	\hspace{-0.0cm}
	\includegraphics[width=0.32\textwidth]{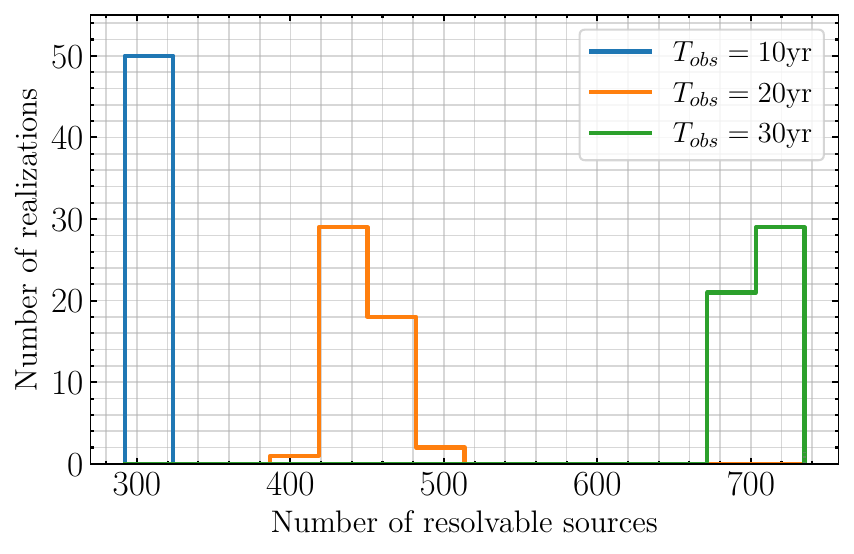}
	\caption{The number of realizations versus the number count of resolvable GW sources for different PTA parameter configurations in the $\tau_{H}=5$ Gyr case. The upper and lower panels represent the results for $N_p=500$ and $1000$, respectively, while the left, middle, and right panels correspond to results for $\sigma_t=200\rm ns$, $100\rm ns$ and $50\rm ns$ respectively.}
	\label{fig:numdistri2}
\end{figure*}

\begin{figure*}
	\centering
	\includegraphics[width=0.32\textwidth]{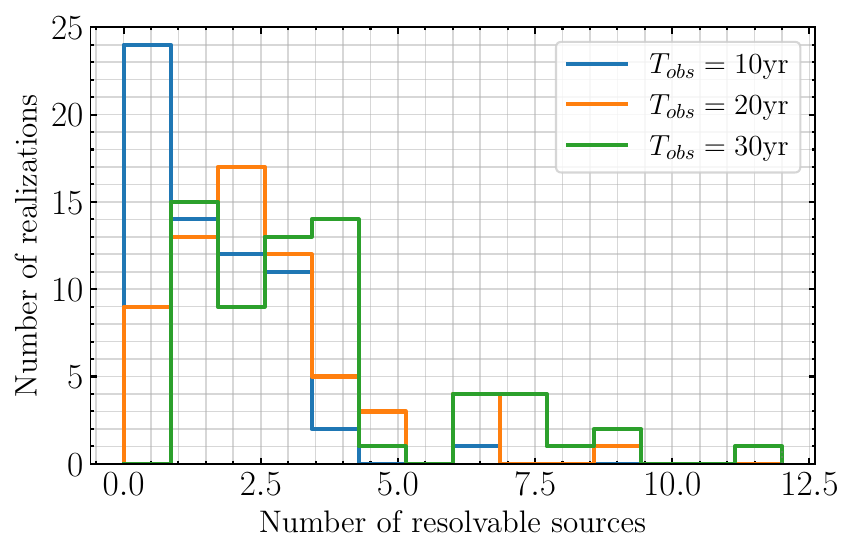}
	\hspace{-0.0cm}
	\includegraphics[width=0.32\textwidth]{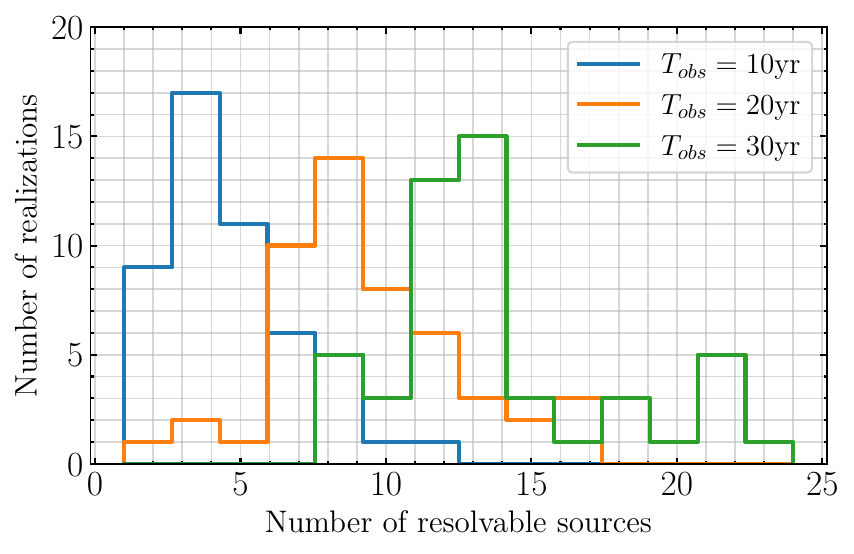}
	\hspace{-0.0cm}
	\includegraphics[width=0.32\textwidth]{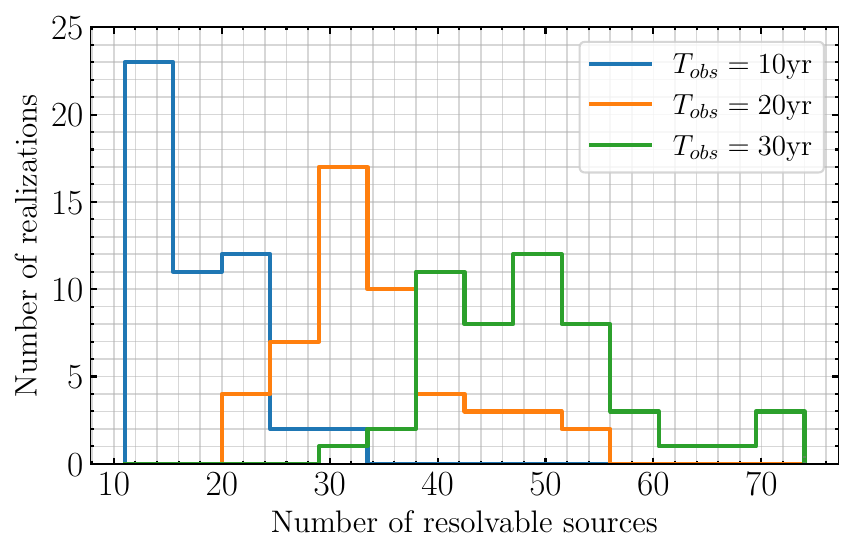}
	\hspace{-0.0cm}
	\includegraphics[width=0.32\textwidth]{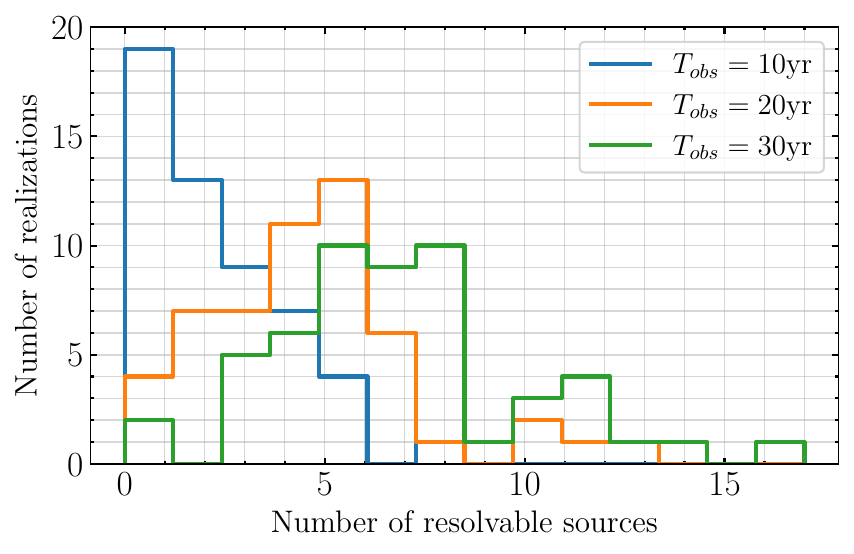}
	\hspace{-0.0cm}
	\includegraphics[width=0.32\textwidth]{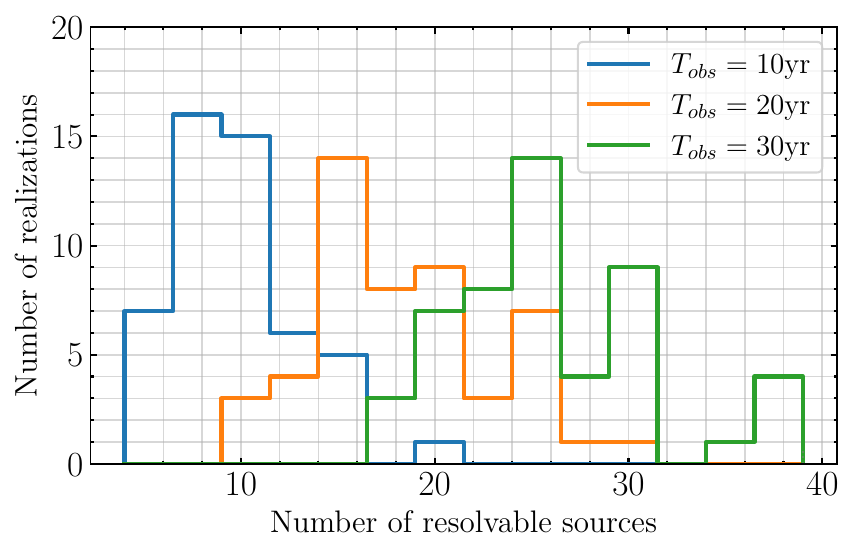}
	\hspace{-0.0cm}
	\includegraphics[width=0.32\textwidth]{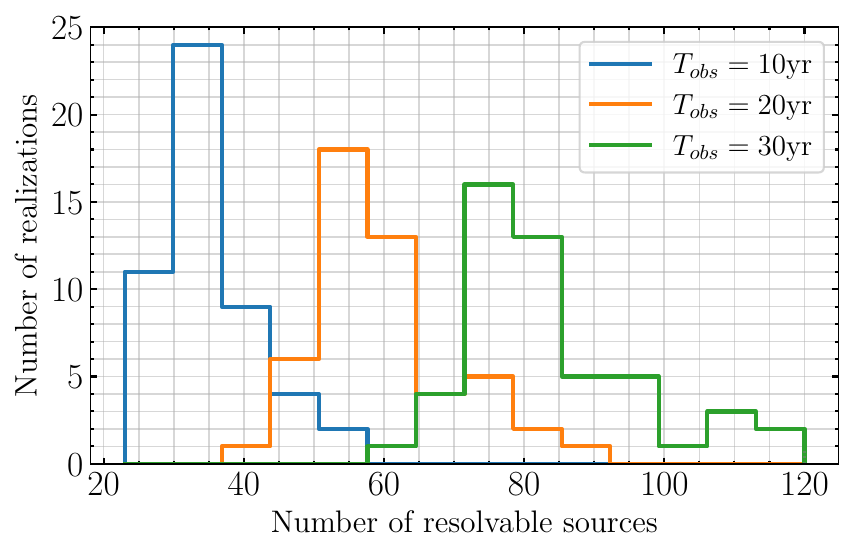}
	\caption{The number of realizations versus the number count of resolvable GW sources for different PTA parameter configurations in the $\tau_{H}=10$ Gyr case. The upper and lower panels represent the results for $N_p=500$ and $1000$, respectively, while the left, middle, and right panels correspond to results for $\sigma_t=200\rm ns$, $100\rm ns$ and $50\rm ns$ respectively.}
	\label{fig:numdistri3}
\end{figure*}

However, it should be noted that not all these bright individual sources can be resolved by PTA detectors. As is given by \cite{2012PhRvD..86l4028B}, due to the limited number of pulsars available to the detector, there is an upper limit to the number of individual sources that can be resolved within each frequency bin. We applied this upper limit of $\sim 2N_p/7$ to each frequency bin, retaining events with higher SNR, and ultimately obtained the dataset of all resolvable individual sources. In Table~\ref{tab:Ndis}, we present the resulted average number of resolvable individual sources (derived from all realizations) for different detector parameter configurations, along with the 1$\sigma$ upper and lower uncertainties for all the three hardening timescales. As shown in Table~\ref{tab:Ndis}, although it's a relatively long time span from 0.1 Gyr to 5 Gyr, the number of resolvable individual sources under both scenarios remains at the same order of magnitude ($10^2$) in the vast majority of cases. However, it is obvious that, above the hardening timescale of 5 Gyr, the number of individual sources decays more rapidly, and even drops to $10^0$ levels at certain parameter configuration in the $\tau_H=10\rm Gyr$ case. For these three hardening timescales, the average numbers of resolvable individual sources under the most pessimistic scenario ($N_p=500$, $\sigma_t=200\rm ns$, $t_{obs}=10\rm yr$) are 111, 30, and 1 for $\tau_H=0.1$, $5$ and $10$ Gyr respectively. While in the most optimistic scenario ($N_p=1000$, $\sigma_t=50\rm ns$, $t_{obs}=30\rm yr$), the average numbers of resolvable individual sources are 706 and 84 for $\tau_H=5$ and $10$ Gyr respectively, and can reach 1590 in the $\tau_H=0.1$ Gyr case. In Fig.~\ref{fig:numdistri1}, \ref{fig:numdistri2}, and \ref{fig:numdistri3}, we also present more detailed distribution histograms showing the number of realizations corresponding to different number counts of resolvable individual sources. 

\begin{table}
	\caption{Tables showing the average number of resolvable sources $n_s$ (with 1$\sigma$ upper and lower uncertainties) under different detector parameters for hardening times of 0.1 Gyr (top panel), 5 Gyr (middle panel), and 10 Gyr (bottom panel).
	}
	\begin{center}
		\begin{tabular}{c|ccc|ccc}
			\multicolumn{7}{c}{$\tau_{H}=0.1 \rm Gyr$} \\
			\hline \hline
			& \multicolumn{3}{c}{$N_p=500$} &  \multicolumn{3}{c}{$N_p=1000$} \\ \hline
			$\tau_{obs}$/yr & 200ns & 100ns & 50ns & 200ns & 100ns & 50ns\\\hline
			10 & $111_{-10}^{+11}$ & $149_{-2}^{+3}$ & $171_{-5}^{+6}$ & $254_{-15}^{+16}$ & $299_{-4}^{+4}$ & $344_{-9}^{+9}$\\ 
			20 & $156_{-4}^{+5}$  &  $207_{-9}^{+9}$ & $342_{-7}^{+8}$ & $315_{-6}^{+7}$ & $420_{-11}^{+13}$ & $828_{-108}^{+123}$\\
			30 & $187_{-5}^{+7}$ & $338_{-6}^{+7}$ & $507_{-14}^{+14}$ & $386_{-9}^{+11}$ & $683_{-11}^{+13}$ & $1590_{-22}^{+26}$\\ \hline \hline
		\end{tabular}
		%
		\vspace{0.3cm}
		
		\begin{tabular}{c|ccc|ccc}
			\multicolumn{7}{c}{$\tau_{H}=5 \rm Gyr$} \\
			\hline \hline
			& \multicolumn{3}{c}{$N_p=500$} &  \multicolumn{3}{c}{$N_p=1000$} \\ \hline
			$\tau_{obs}$/yr & 200ns & 100ns & 50ns & 200ns & 100ns & 50ns\\\hline
			10 & $30_{-7}^{+6}$ & $143_{-6}^{+3}$ & $152_{-4}^{+5}$ & $68_{-8}^{+8}$ & $290_{-5}^{+3}$ & $304_{-6}^{+6}$\\ 
			20 & $67_{-8}^{+9}$  &  $163_{-5}^{+4}$ & $225_{-10}^{+12}$  &  $148_{-13}^{+15}$ & $327_{-6}^{+7}$ & $448_{-15}^{+20}$\\
			30 & $107_{-10}^{+11}$ & $204_{-10}^{+9}$ & $353_{-13}^{+11}$ & $234_{-11}^{+16}$ & $415_{-15}^{+15}$ & $706_{-14}^{+14}$\\ \hline \hline
		\end{tabular}
		
		\vspace{0.3cm}  

		\begin{tabular}{c|ccc|ccc}
			\multicolumn{7}{c}{$\tau_{H}=10 \rm Gyr$} \\
			\hline \hline
			& \multicolumn{3}{c}{$N_p=500$} &  \multicolumn{3}{c}{$N_p=1000$} \\ \hline
			$\tau_{obs}$/yr & 200ns & 100ns & 50ns & 200ns & 100ns & 50ns\\\hline
			10 & $1_{-1}^{+2}$ & $5_{-2}^{+3}$ & $17_{-4}^{+6}$ & $2_{-1}^{+2}$ & $9_{-2}^{+4}$ & $34_{-5}^{+8}$\\ 
			20 & $2_{-1}^{+2}$  &  $9_{-3}^{+4}$ & $34_{-7}^{+9}$  &  $5_{-2}^{+3}$ & $18_{-4}^{+6}$ & $60_{-7}^{+13}$\\
			30 & $3_{-2}^{+3}$ & $14_{-3}^{+5}$ & $49_{-8}^{+10}$ & $7_{-3}^{+3}$ & $26_{-5}^{+6}$ & $84_{-9}^{+17}$\\ \hline \hline
		\end{tabular}
		
	\end{center}
	\label{tab:Ndis}
\end{table}

\subsection{Constraints on Dark Energy EoS Parameter}
\begin{table*}
	\caption{The average measurement errors $\Delta w$ on $w$ (with 1$\sigma$ upper and lower uncertainties) under different detector parameter configurations for hardening timescales of 0.1 Gyr (top panel), 5 Gyr (middle panel) and 10 Gyr (bottom panel) when all the resolvable sources exhibit observable electromagnetic counterparts.
	}
	\begin{center}
		\begin{tabular}{c|ccc|ccc}
			\multicolumn{7}{c}{$\tau_{H}=0.1 \rm Gyr$} \\
			\hline \hline
			& \multicolumn{3}{c}{$N_p=500$} &  \multicolumn{3}{c}{$N_p=1000$} \\ \hline
			$\tau_{obs}$/yr & 200ns & 100ns & 50ns & 200ns & 100ns & 50ns\\\hline
			10 & $0.225_{-0.024}^{+0.028}$ & $0.120_{-0.010}^{+0.011}$ & $0.068_{-0.008}^{+0.007}$ & $0.124_{-0.008}^{+0.010}$ & $0.072_{-0.005}^{+0.006}$ & $0.042_{-0.003}^{+0.004}$\\ 
			20 & $0.149_{-0.016}^{+0.011}$  &  $0.086_{-0.007}^{+0.008}$ & $0.048_{-0.005}^{+0.005}$ & $0.092_{-0.006}^{+0.005}$ & $0.052_{-0.003}^{+0.003}$ & $0.029_{-0.002}^{+0.002}$\\
			30 & $0.123_{-0.010}^{+0.010}$ & $0.066_{-0.005}^{+0.005}$ & $0.037_{-0.003}^{+0.003}$ & $0.073_{-0.004}^{+0.005}$ & $0.040_{-0.002}^{+0.002}$ & $0.023_{-0.001}^{+0.001}$\\ \hline \hline
		\end{tabular}
		%
		
		\vspace{0.3cm}
		
		\begin{tabular}{c|ccc|ccc}
			\multicolumn{7}{c}{$\tau_{H}=5 \rm Gyr$} \\
			\hline \hline
			& \multicolumn{3}{c}{$N_p=500$} &  \multicolumn{3}{c}{$N_p=1000$} \\ \hline
			$\tau_{obs}$/yr & 200ns & 100ns & 50ns & 200ns & 100ns & 50ns\\\hline
			10 & $0.777_{-0.163}^{+0.242}$ & $0.242_{-0.030}^{+0.025}$ & $0.141_{-0.016}^{+0.017}$ & $0.415_{-0.073}^{+0.116}$ & $0.145_{-0.011}^{+0.011}$ & $0.088_{-0.009}^{+0.009}$\\ 
			20 & $0.405_{-0.066}^{+0.074}$  &  $0.177_{-0.021}^{+0.015}$ & $0.098_{-0.013}^{+0.012}$  &  $0.235_{-0.029}^{+0.033}$ & $0.110_{-0.009}^{+0.008}$ & $0.061_{-0.005}^{+0.004}$\\
			30 & $0.290_{-0.029}^{+0.042}$ & $0.145_{-0.021}^{+0.019}$ & $0.078_{-0.009}^{+0.009}$ & $0.172_{-0.017}^{+0.016}$ & $0.087_{-0.007}^{+0.008}$ & $0.048_{-0.004}^{+0.004}$\\ \hline \hline
		\end{tabular}
		
		\vspace{0.3cm}  

		\begin{tabular}{c|ccc|ccc}
			\multicolumn{7}{c}{$\tau_{H}=10 \rm Gyr$} \\
			\hline \hline
			& \multicolumn{3}{c}{$N_p=500$} &  \multicolumn{3}{c}{$N_p=1000$} \\ \hline
			$\tau_{obs}$/yr & 200ns & 100ns & 50ns & 200ns & 100ns & 50ns\\\hline
			10 & $24.726_{-13.907}^{+81.258}$ & $6.974_{-3.120}^{+9.253}$ & $1.730_{-0.513}^{+0.685}$ & $38.481_{-27.638}^{+517.720}$ & $2.956_{-0.895}^{+1.673}$ & $1.133_{-0.235}^{+0.223}$\\ 
			20 & $15.926_{-8.815}^{+32.884}$  &  $3.487_{-1.076}^{+2.086}$ & $1.105_{-0.275}^{+0.295}$  &  $7.854_{-3.906}^{+13.710}$ & $1.879_{-0.410}^{+1.110}$ & $0.715_{-0.127}^{+0.137}$\\
			30 & $12.860_{-7.476}^{+22.071}$ & $2.246_{-0.584}^{+1.086}$ & $0.860_{-0.149}^{+0.219}$ & $4.865_{-2.146}^{+9.367}$ & $1.364_{-0.323}^{+0.395}$ & $0.569_{-0.100}^{+0.094}$\\ \hline \hline
		\end{tabular}
		
	\end{center}
	\label{tab:deltawdis}
\end{table*}

\begin{table*}
	\caption{The average measurement errors $\Delta w$ on $w$ (with 1$\sigma$ upper and lower uncertainties) under different detector parameter configurations for hardening timescales of 0.1 Gyr (top panel) and 5 Gyr (bottom panel) when 10\% of the resolvable sources exhibit observable electromagnetic counterparts.
	}
	\begin{center}
		\begin{tabular}{c|ccc|ccc}
			\multicolumn{7}{c}{$\tau_{H}=0.1 \rm Gyr$} \\
			\hline \hline
			& \multicolumn{3}{c}{$N_p=500$} &  \multicolumn{3}{c}{$N_p=1000$} \\ \hline
			$\tau_{obs}$/yr & 200ns & 100ns & 50ns & 200ns & 100ns & 50ns\\\hline
			10 & $0.830_{-0.218}^{+0.327}$ & $0.443_{-0.113}^{+0.163}$ & $0.259_{-0.073}^{+0.114}$ & $0.420_{-0.076}^{+0.096}$ & $0.245_{-0.047}^{+0.070}$ & $0.143_{-0.029}^{+0.037}$\\ 
			20 & $0.525_{-0.118}^{+0.187}$  &  $0.303_{-0.067}^{+0.101}$ & $0.167_{-0.037}^{+0.052}$ & $0.310_{-0.055}^{+0.067}$ & $0.177_{-0.034}^{+0.041}$ & $0.097_{-0.015}^{+0.018}$\\
			30 & $0.429_{-0.098}^{+0.126}$ & $0.223_{-0.046}^{+0.061}$ & $0.124_{-0.024}^{+0.035}$ & $0.242_{-0.040}^{+0.050}$ & $0.133_{-0.021}^{+0.023}$ & $0.075_{-0.011}^{+0.014}$\\ \hline \hline
		\end{tabular}
		%
		
		\vspace{0.3cm}
		
		\begin{tabular}{c|ccc|ccc}
			\multicolumn{7}{c}{$\tau_{H}=5 \rm Gyr$} \\
			\hline \hline
			& \multicolumn{3}{c}{$N_p=500$} &  \multicolumn{3}{c}{$N_p=1000$} \\ \hline
			$\tau_{obs}$/yr & 200ns & 100ns & 50ns & 200ns & 100ns & 50ns\\\hline
			10 & $5.584_{-2.939}^{+14.241}$ & $0.874_{-0.206}^{+0.277}$ & $0.529_{-0.152}^{+0.228}$ & $1.767_{-0.608}^{+1.104}$ & $0.508_{-0.100}^{+0.125}$ & $0.313_{-0.071}^{+0.085}$\\ 
			20 & $1.819_{-0.643}^{+1.617}$  &  $0.654_{-0.169}^{+0.221}$ & $0.368_{-0.100}^{+0.140}$ &  $0.863_{-0.229}^{+0.291}$ & $0.381_{-0.078}^{+0.096}$ & $0.211_{-0.042}^{+0.053}$\\
			30 & $1.117_{-0.324}^{+0.538}$ & $0.530_{-0.132}^{+0.193}$ & $0.281_{-0.064}^{+0.084}$ & $0.607_{-0.143}^{+0.182}$ & $0.296_{-0.061}^{+0.066}$ & $0.162_{-0.031}^{+0.036}$\\ \hline \hline
		\end{tabular}
		
	\end{center}
	\label{tab:deltawdis10}
\end{table*}

\begin{figure*}
	\centering
	\includegraphics[width=0.45\textwidth]{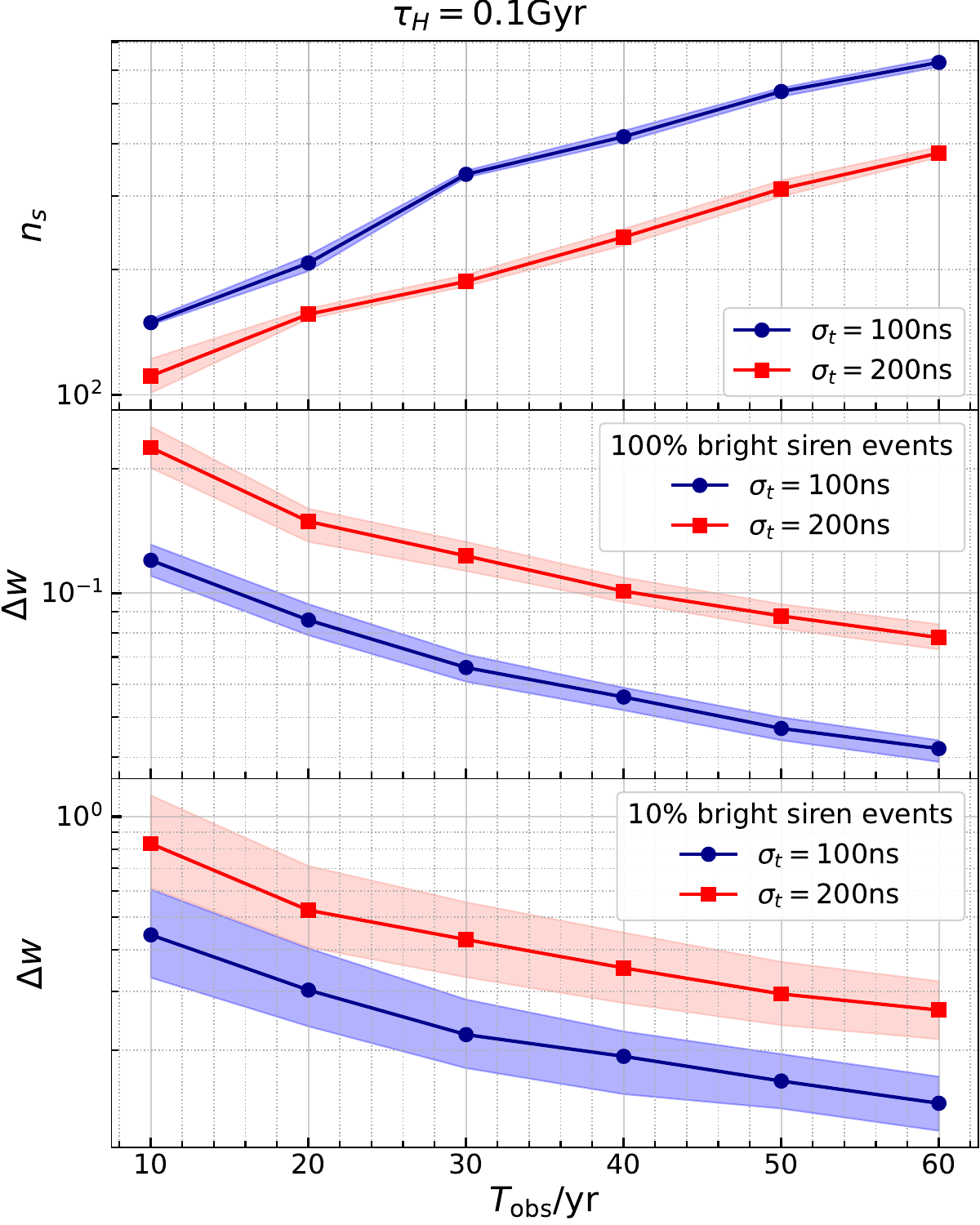}
	\hspace{0.5cm}
	\includegraphics[width=0.45\textwidth]{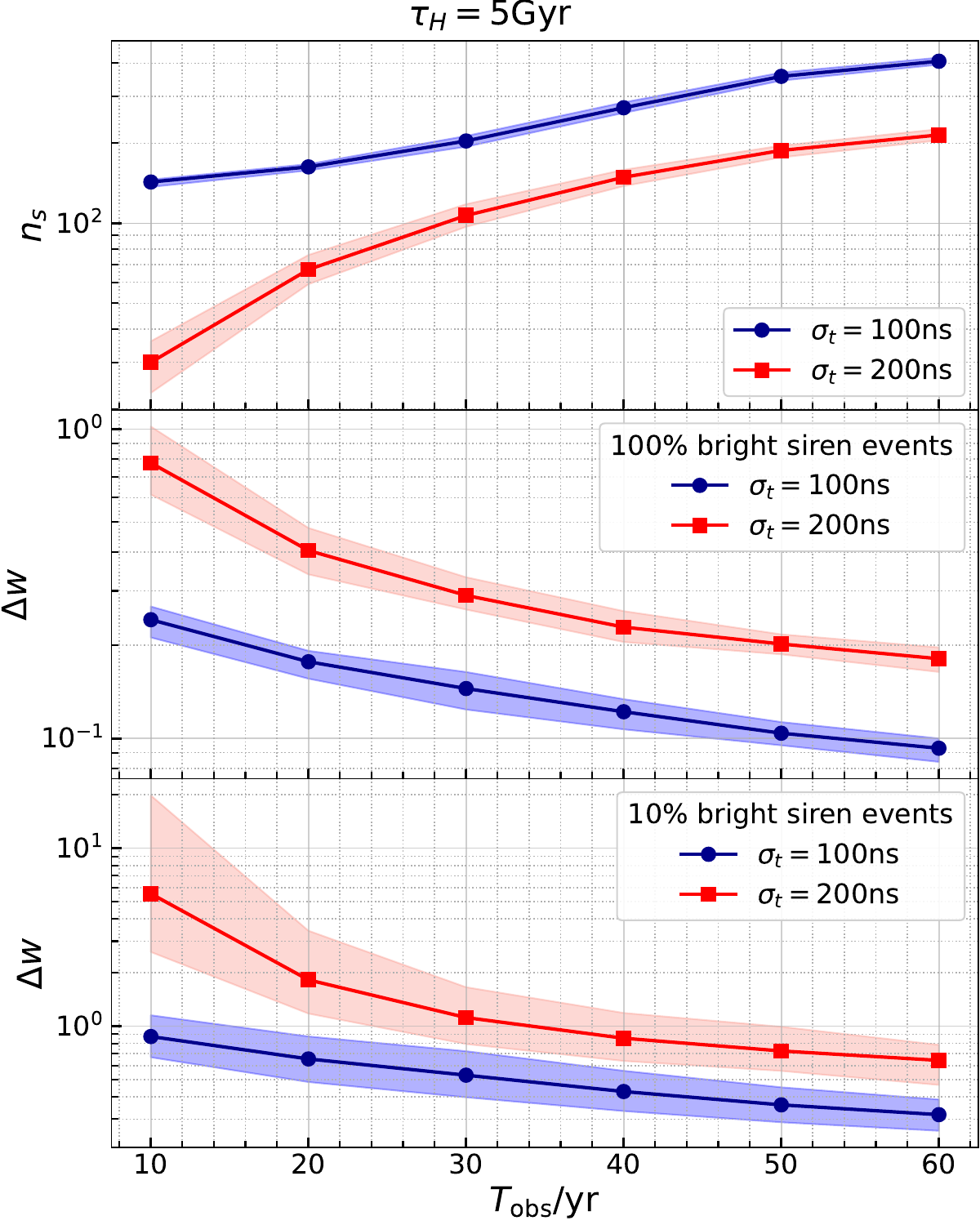}
	\caption{Results under relatively conservative PTA detector configurations (500 pulsars with timing error of 100/200 ns) showing: (upper panels) the number of resolvable individual sources $n_s$, and (middle/lower panels) measurement errors $\Delta w$ on the dark energy EoS parameter $w$ with longer-term data accumulation. Columns correspond to hardening timescales $\tau_H=0.1$Gyr (left) and $\tau_H=5$Gyr (right). Middle panels assume 100\% of GW sources are bright sirens, while lower panels consider only 10\% bright siren fractions. The blue or red shaded regions indicate the distribution range within the 1$\sigma$ uncertainty under  corresponding PTA  parameters.}
	\label{fig:multiyears}
\end{figure*}

\begin{figure*}
	\centering
	\includegraphics[width=0.32\textwidth]{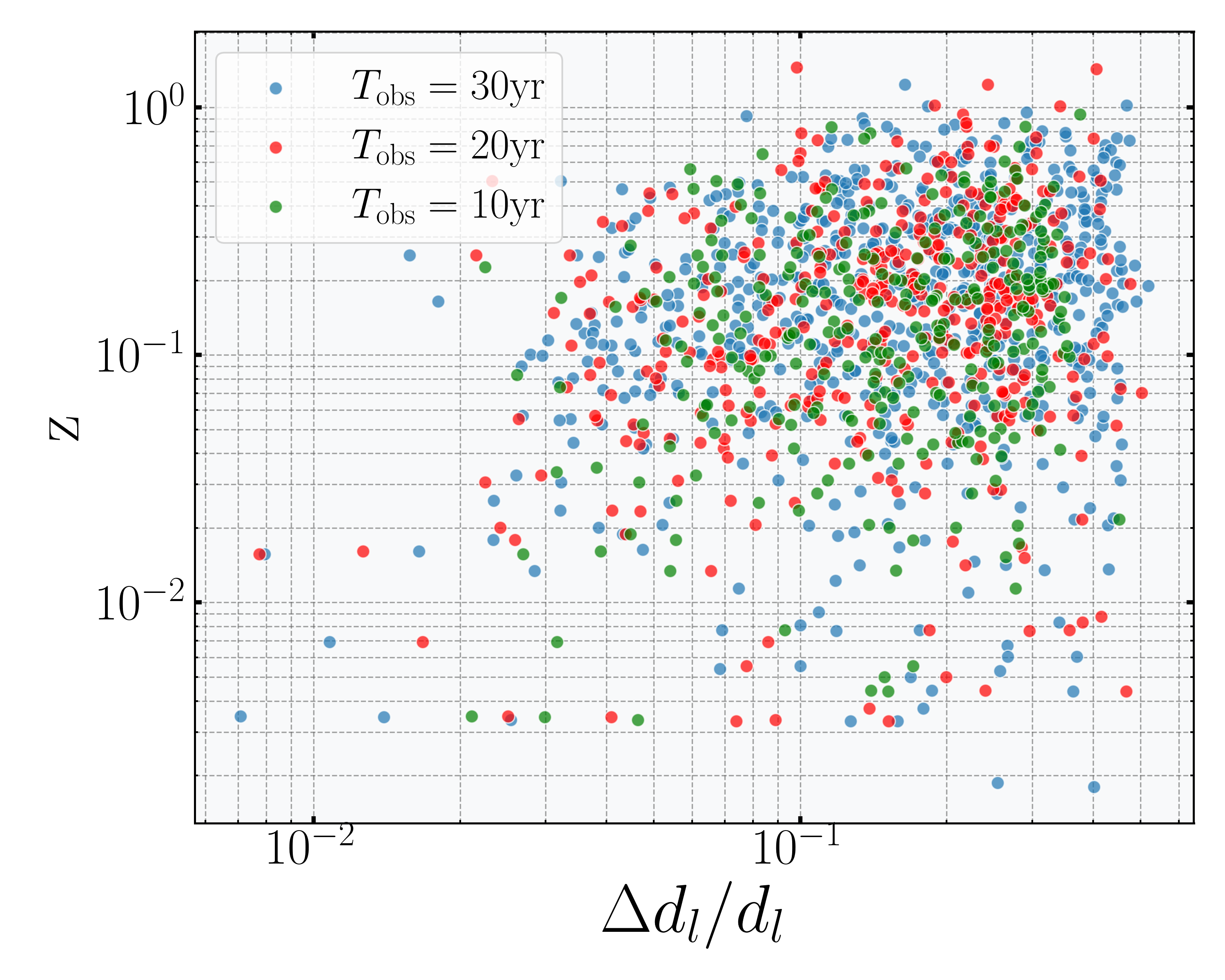}
	\hspace{-0.0cm}
	\includegraphics[width=0.32\textwidth]{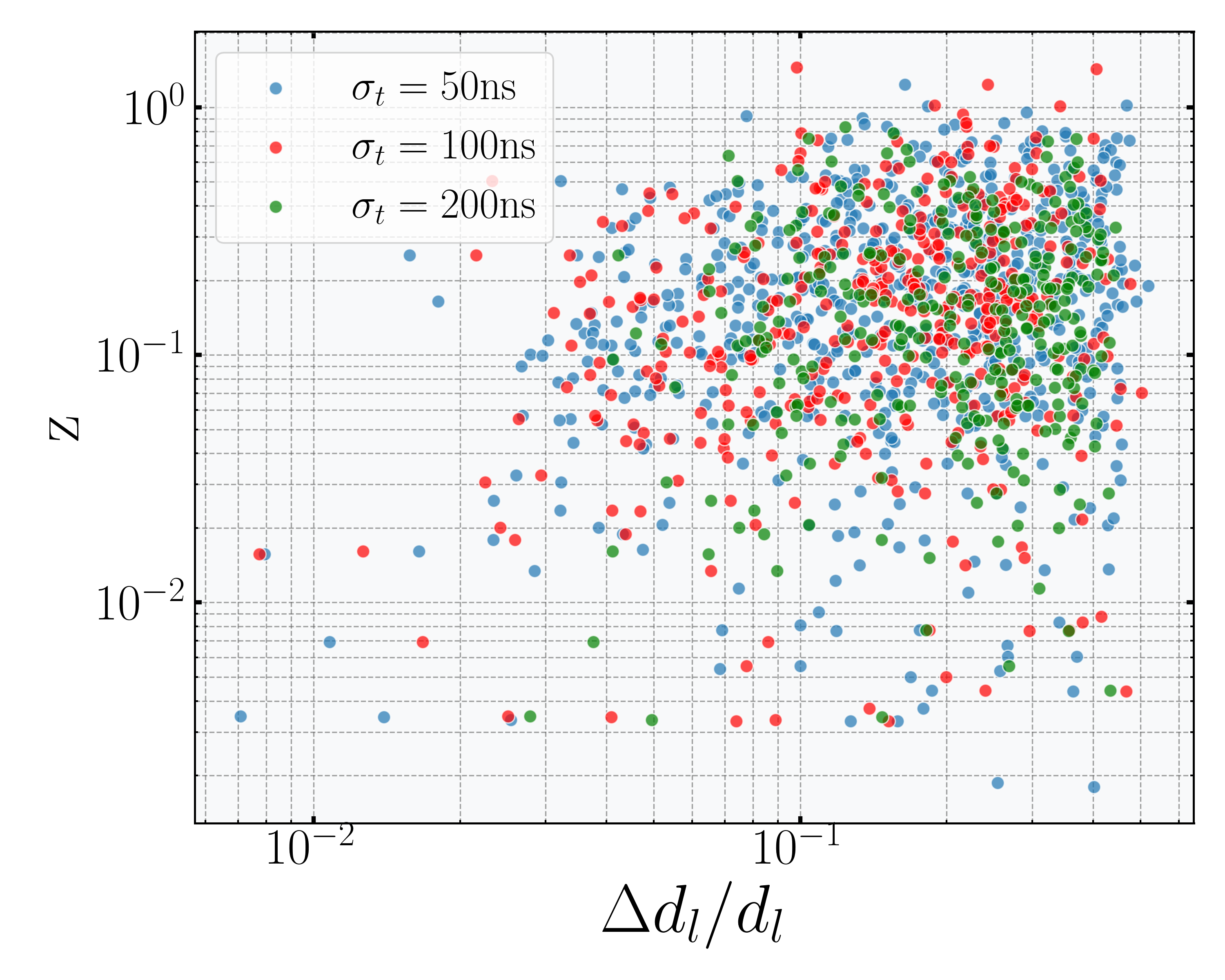}
	\hspace{-0.0cm}
	\includegraphics[width=0.32\textwidth]{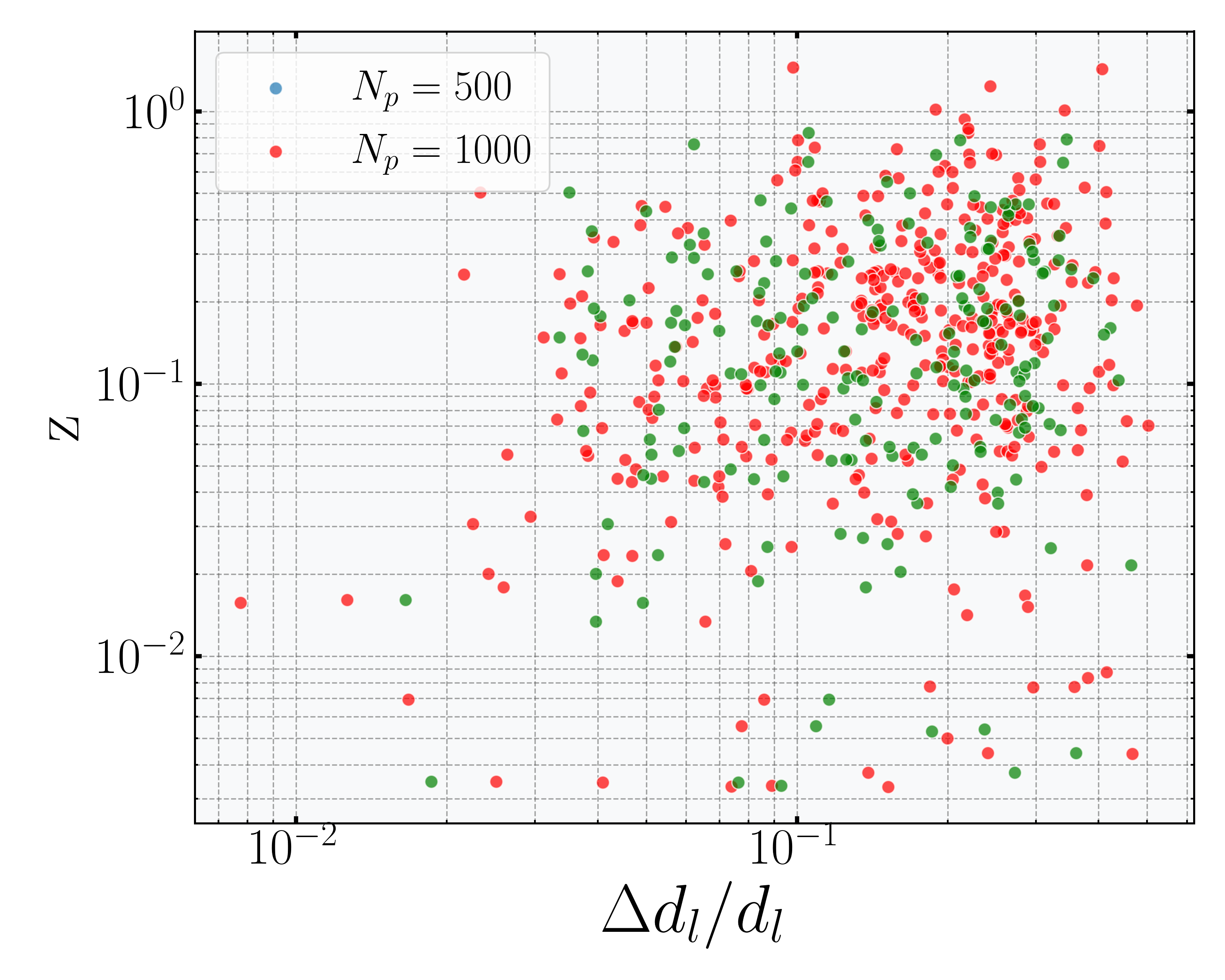}
	\hspace{-0.0cm}
	\includegraphics[width=0.32\textwidth]{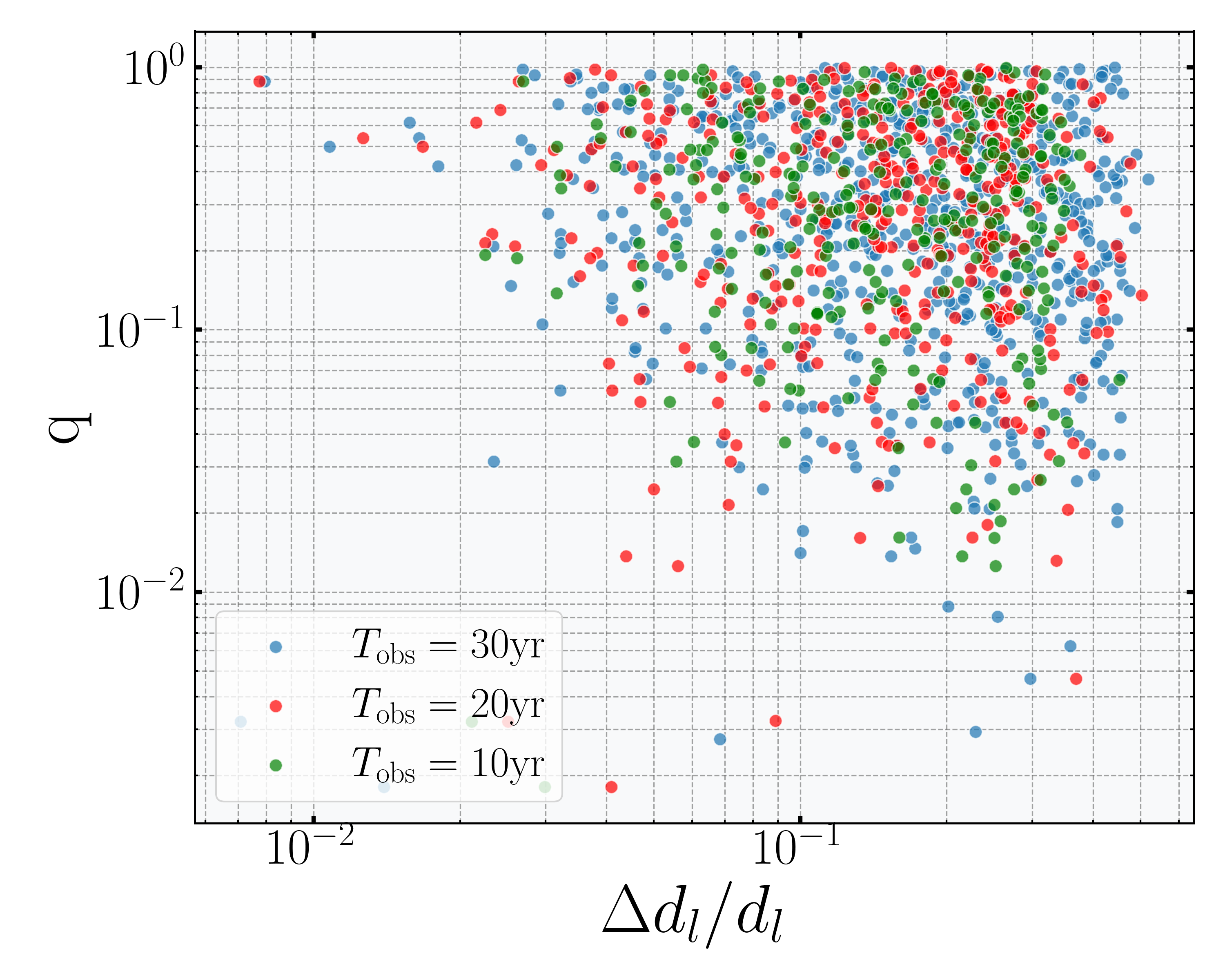}
	\hspace{-0.0cm}
	\includegraphics[width=0.32\textwidth]{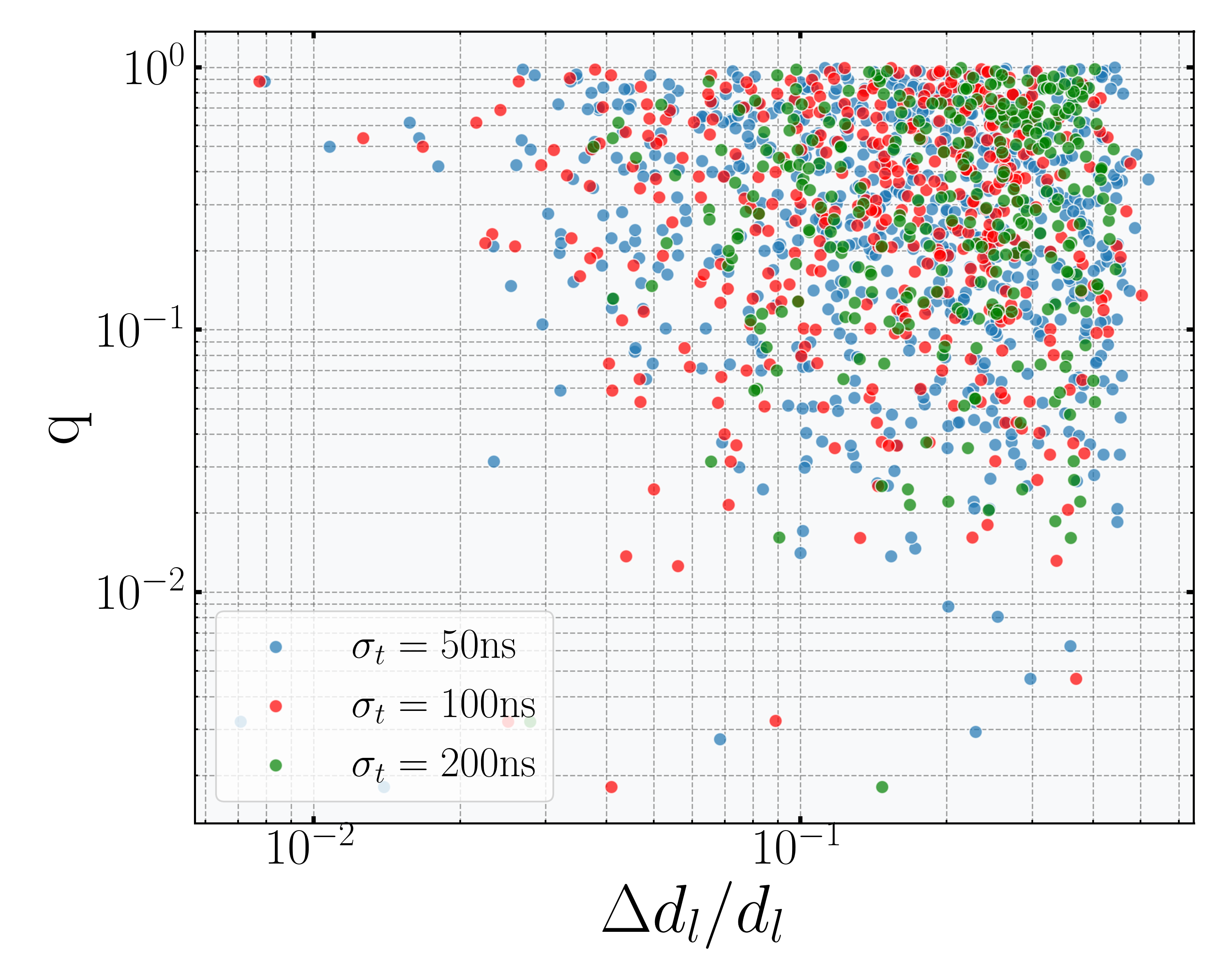}
	\hspace{-0.0cm}
	\includegraphics[width=0.32\textwidth]{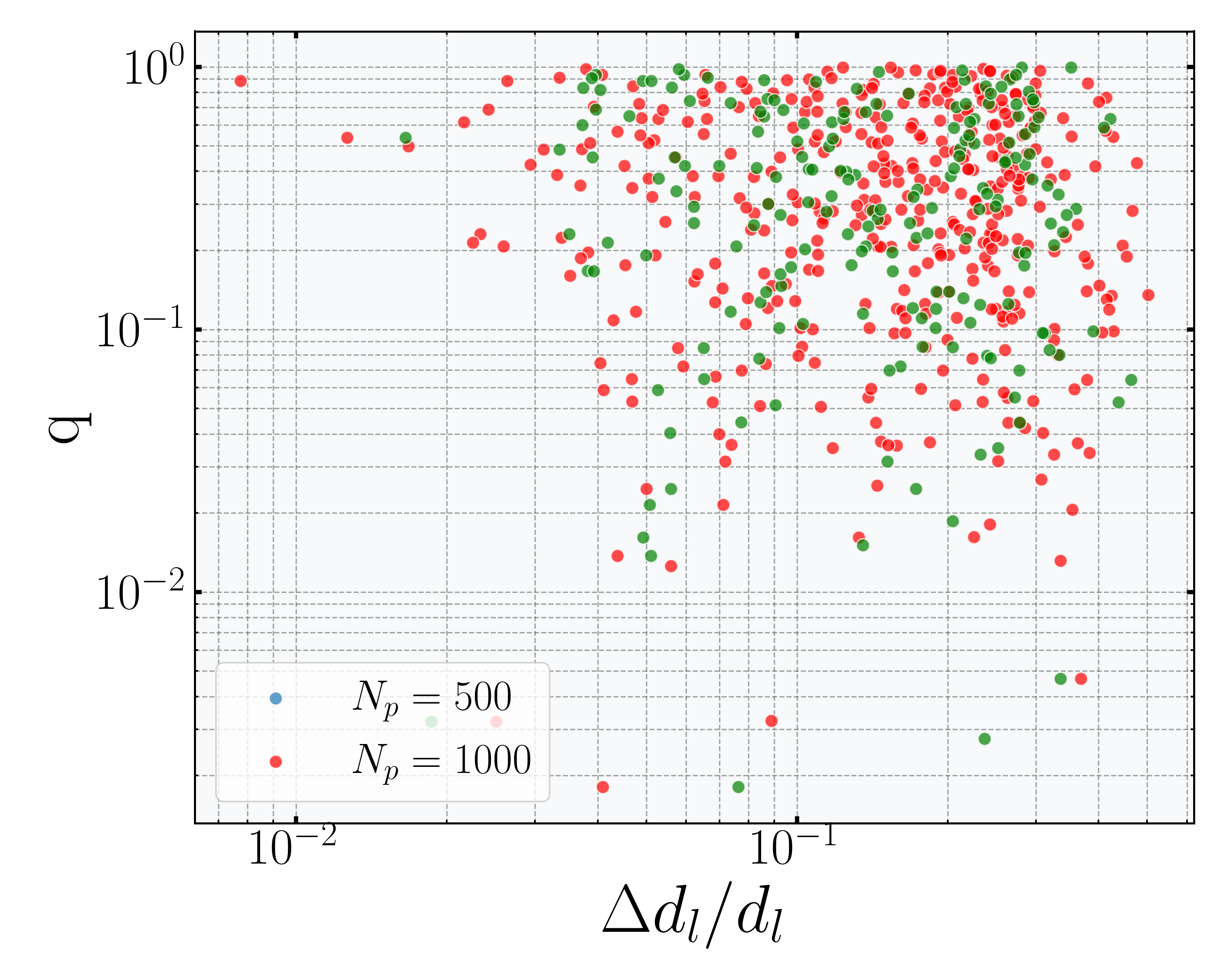}
	\hspace{-0.0cm}
	\includegraphics[width=0.32\textwidth]{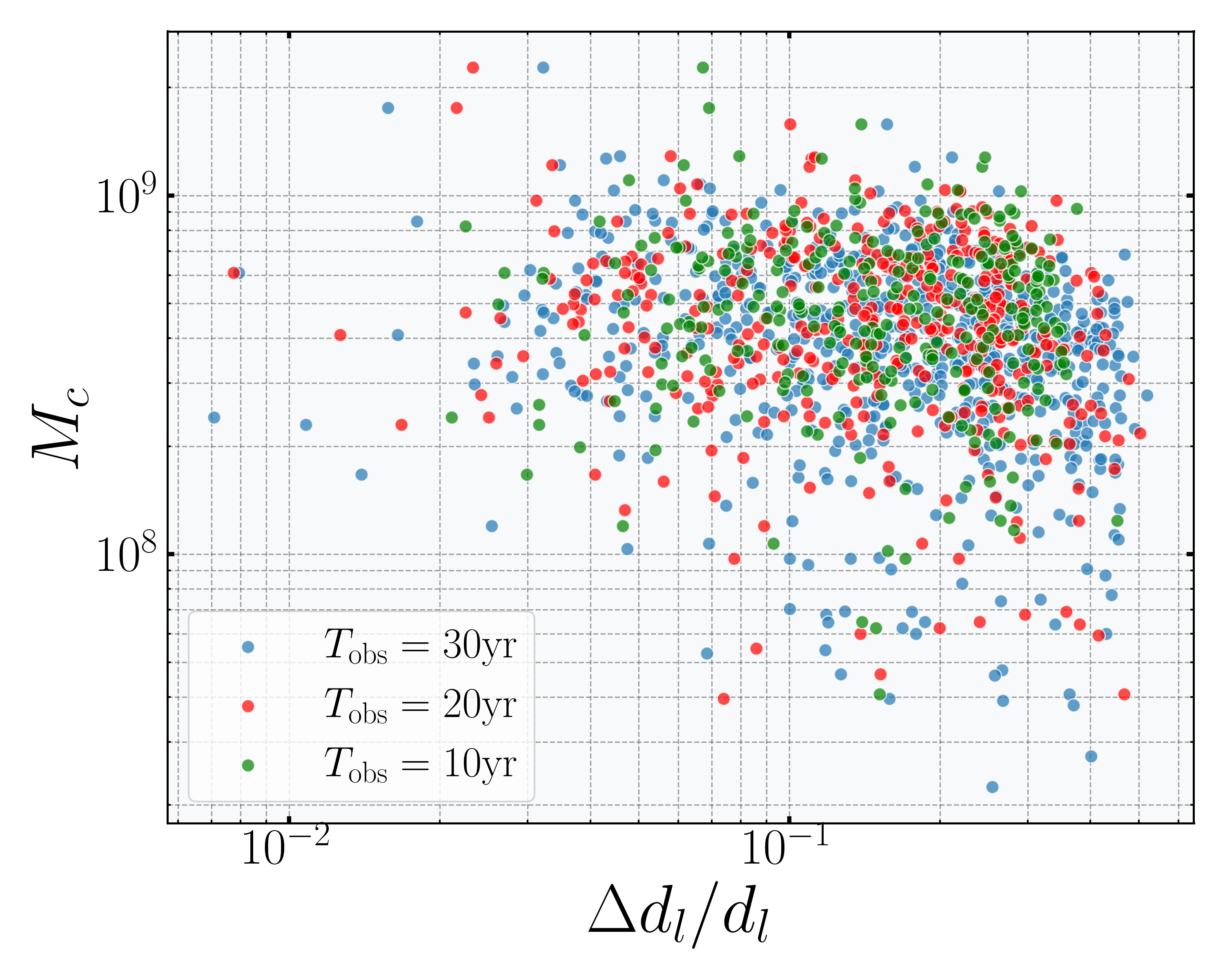}
	\hspace{-0.0cm}
	\includegraphics[width=0.32\textwidth]{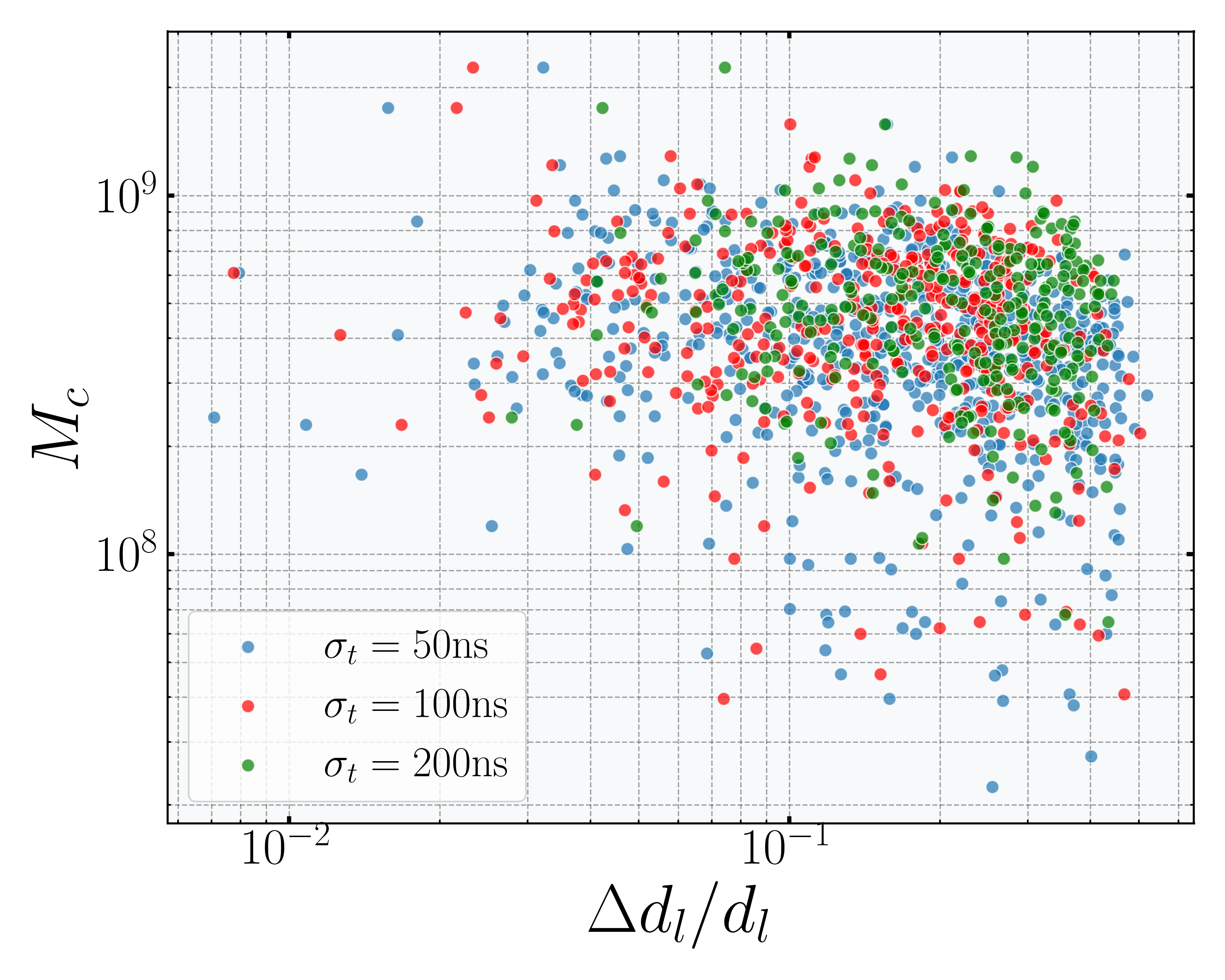}
	\hspace{-0.0cm}
	\includegraphics[width=0.32\textwidth]{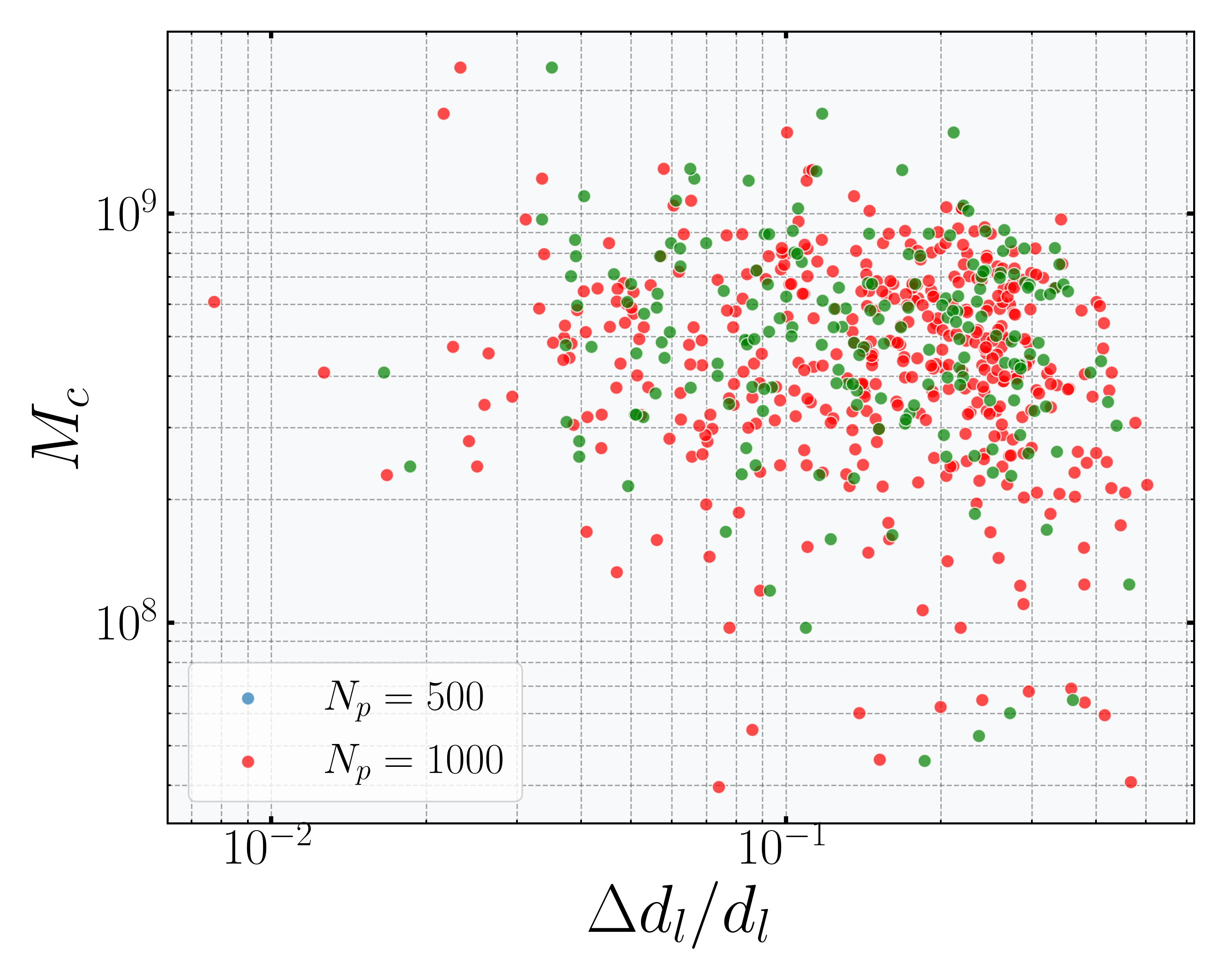}
	\caption{Scatter plots showing the distributions of relative luminosity distance measurement error $\Delta d_l/d_l$ versus redshift (top row), mass ratio of binary black holes (middle row), and chirp mass (bottom row) for all bright sources with SNR $>$ 8 in the most optimal case with hardening timescale $\tau_H=0.1\rm Gyr$ and assuming all events are associated with observable electromagnetic counterparts. The left column displays results for different observation duration (with other parameters set as $Np=1000$, $\sigma_t$=100 ns); The middle column displays results for different timing error level (with other parameters set as $Np=1000$, $T_{\rm obs}=20\rm yr$); The right column displays results for different numbers of pulsars (with other parameters set as $\sigma_t$=100 ns, $T_{\rm obs}=20\rm yr$).}
	\label{fig:scatterdl}
\end{figure*}

Using Equation~\eqref{eq:delta_w_single}, we calculate the measurement error of the dark energy EoS parameter $w$ for each resolvable gravitational wave event, and employ Equation~\eqref{eq:delta_w_combined} to determine the combined constraint from all individual sources. 

In the first scenario where all resolvable individual sources are assumed to have detectable electromagnetic counterparts,  the resulted average measurement errors from all realizations along with the 1$\sigma$ upper and lower uncertainties under three hardening timescales and various PTA  parameter configurations are all presented in Table~\ref{tab:deltawdis}. As clearly shown in the table, the detection sensitivity to dark energy decreases with increasing hardening timescales. This trend naturally aligns with expectations, as longer hardening times lead to fewer detectable individual sources, thereby reducing the measurement precision. It can be observed that for hardening times of 0.1 Gyr, 5 Gyr, and 10 Gyr, the measurement uncertainties on $w$ lie approximately at the order of 0.01, 0.1, and 0.1–1, respectively, for the majority of parameter configurations. Despite the nearly two-order-of-magnitude difference in hardening times between 0.1 Gyr and 5 Gyr, the measurement errors on $w$ remain relatively comparable under both scenarios, indicating a weaker-than-expected dependence of the dark energy constraint power on the hardening timescale within this range. In the most optimistic scenario—with timing residuals of 50 ns and a PTA consisting of 1000 pulsars accumulating 30 years of observations—the measurement errors on the dark energy parameter can reach levels of 0.023 and 0.048 under hardening timescales of 0.1 Gyr and 5 Gyr, respectively. This demonstrates that even for hardening timescales reaching 5 Gyr, gravitational wave detection remains a viable approach for probing parameter $w$ in this scenario, provided substantial improvements in PTA performance (e.g., reduced timing errors) and increases in the number of monitored pulsars are achieved. Moreover, among these factors, enhancements in timing accuracy contribute more significantly to improving the measurement precision of the parameter $w$, as higher timing accuracy directly enhances the detectability of gravitational wave-induced perturbations in pulsar timing datasets. On the other hand, when the hardening timescale increases from 5 Gyr to 10 Gyr, the measurement precision deteriorates significantly. Except in the most optimistic scenario where the measurement uncertainty reaches 0.569, gravitational wave detection of the dark energy parameter $w$ remains infeasible under the majority of parameter configurations. 

For the second scenario where only 10\% of the events are considered to exhibit observable electromagnetic counterpart, the results are presented in Table~\ref{tab:deltawdis10} (Since the hardening timescale $\tau_H=10$Gyr yields unsatisfactory results even under the previously optimistic scenario, we exclude this specific timescale from further consideration in the present scenario). The mean value and the upper/lower uncertainties are derived from statistics of all 500 results, which are obtained by considering 10 different random selections of bright siren events for each of the 50 realizations generated by varying Earth locations. It can be seen that the measurement precision for $w$ is significantly degraded in this scenario. In the case with $\tau_H=5$Gyr, the average measurement error for $w$ falls below 0.2 solely under the most favorable conditions. When the hardening time is set to 0.1 Gyr, the results show modest improvement. The average measurement error for $w$ can reach the level below 0.2 under three configurations: either with 500 pulsars achieving a timing precision of 50 ns and over 20 years of accumulated data, or with 1000 pulsars at a timing precision of 100 ns combined with over 20 years of accumulated data, or alternatively using 1000 pulsars with higher timing accuracy (50 ns) and a shorter data span of 10 years. And the measurement error can reach the level of 0.075 under the most favorable conditions. Notably, in the absence of electromagnetic counterparts, statistical cross-matching within the spatial posterior distributions of GW events could probabilistically assign host galaxies, thereby enabling the incorporation of population-level priors. Consequently, the constraints derived from the this scenario could be viewed as upper limits on the measurement error, as we didn't consider potential gains from host galaxy identification here. 

Given the challenges and uncertainties associated with technological upgrades, it is also necessary to investigate the baseline performance of PTAs in resolving individual GW sources and detecting dark energy parameters. In Fig.~\ref{fig:multiyears}, we further present that, under relatively conservative PTA detector configurations (500 pulsars with timing error of 100/200 ns), the projected number of resolvable individual sources and achievable precision for $w$ measurements solely through extended data accumulation (considering up to 60 years of observations). The figure demonstrates consistent behavior across different parameter configurations: with an additional 30 years of data accumulation, the number of resolvable individual sources approximately doubles, while the measurement precision for $w$ improves by approximately 40\%.


Finally, since the measurement errors on $w$ result directly from the uncertainties in the determination of luminosity distance. In Fig.~\ref{fig:scatterdl} we also show the scatter plots showing the distributions of relative measurement error of luminorsity distance versus other parameters like redshift, mass ratio, or chirp mass for all resolvable sources under different PTA detector parameters. These distributions are derived from the most optimal case with $\tau_H=0.1\rm Gyr$ for the first scenario assuming all events are associated with observable electromagnetic counterparts. For the majority of events, the relative measurement errors of luminosity distance range from $10^{-2}$ to $5\times10^{-1}$. It is evident that events with lower redshift, higher mass ratios or larger chirp masses tend to achieve higher precision in luminosity distance measurements, which of course aligns with theoretical expectations, although the dependence of luminosity distance measurement precision on chirp mass appears less pronounced compared to the former two factors.
\section{Conclusions and Discussions}
\label{sec:concl}

This study explored the detectability of individual SMBBH systems in the nanohertz GW band and their potential to constrain dark energy parameters using next-generation PTAs. By leveraging cosmological simulations based on Millennium database and semi-analytic galaxy formation models, we generated multiple realizations of light-cone SMBBHs across varying hardening timescales ($\tau_H = 0.1, 5$ and $10$ Gyr). We studied the distribution of the SNR of the GW sources and selected those sources that are resolvable with SNR$>$8 under different PTA parameter configurations. Our analysis revealed that advanced PTAs, such as the SKA-era arrays with improved timing precision, extended observational baselines, and larger pulsar networks, could resolve hundreds to thousands of individual SMBBH sources. These sources could offer a pathway to constrain the dark energy EoS parameter $w$ with uncertainties as low as $\Delta w \sim $ 0.023–0.048 for favorable astrophysical scenarios ($\tau_H < 5$ Gyr) when $\sigma_t = 50$ ns, $T_{obs} = 30 \rm yr$, $N_p = 1000$, and assuming all resolvable individual sources have detectable electromagnetic counterparts. Furthermore, when $\tau_H$ is 0.1 Gyr, the measurement uncertainties on $w$ can reach the order of 0.01 for the majority of PTA parameter configurations. In the $\tau_H=5$ Gyr case, while correspondingly larger to some extent, the measurement errors on $w$ are relatively comparable to that in the $\tau_H=0.1$ Gyr case, indicating a weaker-than-expected dependence of the dark energy constraint precision on the hardening timescale within this range.
We also considered a more conservative scenario where only 10\% of gravitational wave sources have detectable electromagnetic counterparts. In this scenario, for $\tau_H = 5~\text{Gyr}$, the average uncertainty in parameter $w$ remains $\Delta w < 0.2$ only under optimal detector configurations (e.g., $N_p = 1000$, $\sigma_t = 50~\text{ns}$, $T_{\text{obs}} = 30~\text{yr}$). The $\tau_H = 0.1$ Gyr case shows enhanced precision: sub-0.2 uncertainties emerge in three scenarios: (1) $N_p = 500$ with $\sigma_t = 50~\text{ns}$ and $T_{\text{obs}} \geq 20~\text{yr}$, (2) $N_p = 1000$ with $\sigma_t = 100~\text{ns}$ and $T_{\text{obs}} \geq 20~\text{yr}$, or (3) $N_p = 1000$ with $\sigma_t = 50~\text{ns}$ even at $T_{\text{obs}} = 10~\text{yr}$. The most favorable configuration ($N_p = 1000$, $\sigma_t = 50~\text{ns}$, $T_{\text{obs}} = 30~\text{yr}$) achieves $\Delta w \sim 0.075$. Furthermore, we investigated the projected number of resolvable individual sources and achievable precision for $w$ measurements solely through extended data accumulation under relatively conservative PTA detector configurations (500 pulsars with timing error of 100/200 ns), and found that with an additional 30 years of data accumulation, the number of resolvable individual sources approximately doubles, while the measurement precision for $w$ improves by approximately 40\%.


Now we offer some further discussions. Several simplifications in our framework warrant refinement in future studies. First, our analysis assumed circular orbits for all binary systems, neglecting potential eccentricity effects \cite{2025ApJ...978..104G}. Eccentricity could alter GW emission patterns, orbital decay rates, and parameter estimation accuracy, particularly for dynamically formed binaries in dense galactic nuclei. Incorporating eccentric orbital evolution would refine predictions for GW strain amplitudes and merger timescales. Second, We considered two scenarios: (1) an optimistic case where all the sources have electromagnetic counterparts, and (2) a conservative case where only 10\% of sources exhibit detectable electromagnetic counterparts. In practice, the detection of electromagnetic counterparts is inherently fraught with non-trivial uncertainties, obscuration by gas-rich environments, GW event localization uncertainty or other technical difficulties could limit the possibility for successful detection, specially for high-$z$ sources. We employed random selection process to identify these bright siren events among the GW sources. However, incorporating host galaxy or GW source properties to quantify electromagnetic counterpart detectability may yield improved accuracy. What's more, statistical cross-matching
within the spatial posterior distributions of the dark siren GW events may result in enhanced detection precision. Thirdly, while realistic PTA noise typically include both white and red noise components from various astrophysical and instrumental sources, this study has focused on an optimistic scenario with only white noise; a more comprehensive treatment considering more realistic noise conditions will be explored in subsequent work. Although such additional noise components may reduce the number of individually resolvable SMBBH sources, lowering the SNR detection threshold could still make dark energy constraints feasible with a population of low-SNR events. Future studies could employ full Bayesian MCMC methods to robustly quantify the precision of dark energy equation-of-state constraints under more realistic noise conditions.
Lastly, in this work, we have only considered a constant form for the dark energy equation of state, i.e., $w$ does not evolve with redshift. However, recent results from the Dark Energy Spectroscopic Instrument (DESI), combined with other probes, mildly favor a dynamical dark energy model with $w_0>-1, w_a<0$ under the Chevallier-Polarski-Linder (CPL) parametrization \cite{DESI:2024mwx,DESI:2025zgx,2025arXiv250915308C,2025arXiv251107517P}. This result shows a slight but intriguing tension with the cosmological constant paradigm, and yet the Hubble tension remains a statistically significant discrepancy, likely requiring pre-recombination new physics for a resolution \cite{Jiang:2024xnu,2024arXiv240418579W,2025arXiv251210585C,2025arXiv251207281Z}. Based on the current statistical significance of the results, it is not yet sufficient to override the broad consistency of $\mathrm{\Lambda CDM}$ with a wide range of cosmological observation, further independent datasets and higher-precision measurements are required to determine whether these tensions reflect new physics or systematic uncertainties. Our methodology can also be extended to dynamical dark energy scenarios by providing joint constraint on the dark energy EoS parameters $w_0$ and $w_a$ in CPL parametrization using either Fisher matrix or MCMC approach, thus enabling a rigorous forecast of the accuracy with which nHz GW signals could serve as an independent probe of dark energy evolution. It is also interesting to note that in modified gravity theories, the behavior of the dark energy EoS parameter $w$ imprints itself on the effective matter-gravity coupling through its impact on the ratio of dark energy to matter perturbations $\delta\rho_{\mathrm{DE}}/\delta\rho_m$, thereby influencing the rate and efficiency of cosmic structure formation\cite{2025arXiv251115049E}. Constraints on dark energy EoS from GW observations, combined with measurements of cosmic structure, may offer independent verification for these models. A nascent alternative approach is to reconstruct the cosmic large scale structure of GW sources in the luminosity distance space \cite{Libanore:2020fim,Palmese:2020kxn,Yang:2022uye,Yang:2024mqz}, which may also serve as an unique channel to constrain effective couplings in modified gravity theories.
Addressing all the issures mentioned above will require tighter integration of numerical simulations and cosmological analyses. We leave these to future works.
\section*{Data Availability}
The data underlying this article are available in Millennium Database at \url{https://gavo.mpa-garching.mpg.de/Millennium/}

\section*{Code Availability}
The code supporting this article can be available on reasonable request.

\section*{Acknowledgements}
We thank Prof. Changshuo Yan for helpful suggestions. Qing Yang is supported by the National Key R\&D Program of China (grant No. 2024YFC2207700), Xiao Guo is supported by the National Natural Science Foundation of China (grant No. 12503001) and the Postdoctoral Fellowship Program and China Postdoctoral Science Foundation (grant No. BX20230104). We acknowledge the Beijing Super Cloud Computing Center for providing HPC resources that have contributed to the research results reported within this paper (URL: \url{http://www.blsc.cn/}).

\bibliographystyle{spmpsci}
\bibliography{ref}
%
%
%

\end{document}